\def\kms{\ifmmode{\,\hbox{km}\,s^{-1}}\else {\rm\,km\,s$^{-1}$}\fi}

\def\kmsm{{\rm\,km\,s^{-1}\,Mpc^{-1}}}

\def\hmpc{\ifmmode{h^{-1}\,\hbox{Mpc}}\else{$h^{-1}$\thinspace Mpc}\fi}
\def\hkpc{\ifmmode{\,h^{-1}\,{\rm kpc}}\else {$h^{-1}$\,kpc}\fi}

\def\etal{{\it et~al.}~}

\def\ie{{\it i.e.}~}

\documentclass{aastex}
\begin{document}

\title{Galaxy Clustering Evolution 
	in the CNOC2 High-Luminosity Sample}

\author{R.~G.~Carlberg\altaffilmark{1,2},
H.~K.~C.~Yee\altaffilmark{1,2},
S.~L.~Morris\altaffilmark{1,3},
H.~Lin\altaffilmark{1,2,4,5}, \\
P.~B.~Hall\altaffilmark{1,2},
D.~Patton\altaffilmark{1,6},
M.~Sawicki\altaffilmark{1,2,7}, 
and
C.~W.~Shepherd\altaffilmark{1,2}
}

\altaffiltext{1}{Visiting Astronomer, Canada--France--Hawaii Telescope, 
        which is operated by the National Research Council of Canada,
        le Centre National de Recherche Scientifique, and the University 
	of Hawaii.}
\altaffiltext{2}{Department of Astronomy, University of Toronto, 
        Toronto ON, M5S~3H8 Canada}
\altaffiltext{3}{Dominion Astrophysical Observatory, 
        Herzberg Institute of Astrophysics,    ,  
        National Research Council of Canada,
        5071 West Saanich Road,
        Victoria, BC, V8X~4M6, Canada}
\altaffiltext{4}{Steward Observatory, University of Arizona,
        Tucson, AZ, 85721}
\altaffiltext{5}{Hubble Fellow}
\altaffiltext{6}{Department of Physics \& Astronomy,
        University of Victoria,
        Victoria, BC, V8W~3P6, Canada}
\altaffiltext{7}{Mail Code 320-47, Caltech, Pasadena 91125, USA}


\begin{abstract} 
The redshift evolution of the galaxy two-point correlation function is
a fundamental cosmological statistic. To identify similar galaxy
populations at different redshifts, we select a strict volume-limited
sample culled from the 6100 cataloged CNOC2 galaxies.  Our
high-luminosity subsample selects galaxies having k-corrected and
evolution-compensated R luminosities, $M_R^{k,e}$, above $-20$ mag
($H_0=100 \kmsm$) where $M_\ast^{k,e}({\rm R})\simeq -20.3$ mag. This
subsample contains about 2300 galaxies distributed between redshifts
0.1 and 0.65 spread over a total of 1.55 square degrees of sky. A
similarly defined low-redshift sample is drawn from the Las Campanas
Redshift Survey.  We find that the co-moving two-point correlation
function can be described as $\xi(r|z) = (r_{00}/r)^\gamma
(1+z)^{-(3+\epsilon-\gamma)}$ with $r_{00}=5.03\pm0.08\hmpc$,
$\epsilon=-0.17\pm 0.18$ and $\gamma=1.87\pm0.07$ over the $z=0.03$ to
0.65 redshift range, for $\Omega_M=0.2, \Lambda=0$. The measured
clustering amplitude and its evolution are dependent on the adopted
cosmology.  The measured evolution rates for $\Omega_M=1$ and flat
$\Omega_M=0.2$ background cosmologies are $\epsilon=0.80\pm0.22$ and
$\epsilon=-0.81\pm0.19$, respectively, with $r_{00}$ of
$5.30\pm0.1\hmpc$ and $4.85\pm0.1\hmpc$, respectively.  The
sensitivity of the derived correlations to the evolution corrections
and details of the measurements is presented.  The analytic prediction
of biased clustering evolution for only the low density, $\Lambda$CDM
cosmology is readily consistent with the observations, with biased
clustering in an open cosmology somewhat marginally excluded and a
biased $\Omega_M=1$ model predicting clustering evolution that is more
than 6 standard deviations from the measured value.
\end{abstract}


\section{Introduction}

The measurement of the evolution of galaxy clustering is a direct test
of theories of the evolution of structure and galaxy formation in the
universe.  Clustering is predicted to change with increasing redshift
in a manner that depends on the background cosmology, the spectrum of
primordial density fluctuations out of which clustering grows, and the
relation of galaxies to dark matter halos.  Specific predictions of
clustering growth are available for a range of CDM-style cosmological
models and galaxy identification algorithms. As a result of biasing
\citep{kaiser84} there is a physically important generic prediction
that the clustering of normal galaxies should not be identical to dark
matter clustering and their clustering should evolve slowly at low
redshifts.  More generally, clustering evolution is of interest for
its impact on galaxies, since clustering leads to galaxy-galaxy
merging and creates the groups and clusters in which galaxies are
subject to high gas densities and temperatures not found in the
general field.

The theoretical groundwork to interpret the quantitative evolution of
dark matter clustering and the trends of galaxy clustering evolution
is largely in place for hierarchical structure models.  Although the
details of the mass buildup of galaxies and the evolution of their
emitted light are far from certain at this time, clustering of
galaxies depends primarily on the distribution of initial density
fluctuations on the mass scale of galaxies.  N-body simulations of
ever growing precision along with their theoretical analysis
\citep{DEFW,virgogal} have led to a good semi-analytic understanding
of dark matter clustering into the nonlinear regime. One result is a
remarkable, convenient, theoretically motivated, empirical equation
that relates the linear power spectrum and its nonlinear outcome 
\citep{EDWF,hamilton,pd}. This allows an analytic prediction
of the clustering evolution of the dark matter density field.

Normal galaxies, which are known to exist near the centers of dark
matter halos with velocity dispersions in the approximate range of 50
to 250 $\kms$, cannot have a clustering evolution identical to the
full dark matter density field. Kaiser (1984) showed that the dense
``peaks'' in the initial density field that ultimately collapse to
form halos are usually more correlated than the full density
field. For high peaks, the peak-peak correlation, $\xi_{\nu\nu}(r|z)$,
is approximately $[\nu/\sigma(z)]^2$ times the correlation of the full
dark matter density field, $\xi_{\rho\rho}(r|z)$, (see Mo \& White
1996 for a more general expression) where $\nu$ measures the minimal
``peak height'' for formation of a halo, in units of the variance on
that mass scale, $\sigma(M,z)$ \citep{bbks}. Both $\sigma^2(M,z)$ and
$\xi_{\rho\rho}(z,r)$ (in co-moving co-ordinates) grow approximately
as $D^2(z,\Omega)$ (exactly so in the linear regime), where
$D(z,\Omega)$ is the growth factor for density perturbations in the
cosmology of interest.  The result is that $\xi_{\nu\nu}(r|z)$ stays
nearly constant in co-moving co-ordinates. This result is
approximately verified in n-body experiments
\citep{cc89,cbg91,ccc,virgo} which can be accurately modeled with a
theoretically motivated analytic function \citep{mowhite}. An
implication is that dark matter halo clustering evolution will have
little sensitivity to the background cosmology
\citep{governato,virgogal}. These results formally apply to low
density ``just virialized'' halos, whereas galaxies are found in the
dense central regions of dark matter halos.  Therefore the clustering
of galaxies needs to take into account their dissipation relative to
the (presumed) dissipationless dark matter.  Dissipationless n-body
simulations which resolve sub-halos within larger virialized units
provide the basic dynamical information but still require a theory of
galaxy formation to associate them with luminous objects, as guided by
observations such as we report here.

At low redshift there have been several substantial clustering
surveys, deriving information from both angular correlations, which
can be de-projected with non-evolving luminosity functions, and
redshift surveys, where the kinematics of the galaxies provides
additional information about clustering dynamics.  The observational
measurement of clustering at higher redshifts has yet to reach the 
size, scale coverage, or redshift precision of the pioneering low
redshift CfA survey \citep{dp}.  Angular correlations of galaxies at
higher redshifts provide some insights, but inevitably mix different
galaxy populations at different redshifts and require an accurate
$n(z)$ to break the degeneracy between an evolving luminosity function
and an evolving clustering amplitude
\citep{ip,widei,csb_hdf}. There are two redshift surveys extending out to
$z\simeq 1$, the Canada France Redshift Survey \citep{cfrs_lf} and the
Hawaii K survey \citep{cowie}.  Measurement of the correlation
evolution of the galaxies in these surveys found a fairly rapid
decline in clustering with redshift \citep{cfrs_xi,kkeck}.  Neither
analysis took into account the evolution of the luminosity function or
was able to quantify the effects of the small sky areas containing the
samples.  Recently two relatively large sky area samples, a
preliminary analysis of the survey reported in this paper
\citep{roysoc} and a shallower limiting magnitude survey \citep{norris}, have
indicated much stronger correlations at about redshift $z\simeq 0.3$
than the earlier small area surveys.

This is the first in a series of papers which discusses the clustering
and kinematic properties of the Canadian Network for Observational
Cosmology field galaxy redshift survey.
\citep{yee2}. The CNOC2 survey is designed to 
be comparable in size and precision to the first CfA redshift survey
\citep{dp} but covering galaxies out to redshift 0.7. In this paper we 
restrict our analysis to measurements of the correlation amplitude of
the high-luminosity galaxies in the CNOC2 redshift survey, which are a
particularly simple and interesting subsample.  The CNOC2 luminosity
function is known in considerable detail \citep{huan_lf} which allows
us to define an evolution-compensated volume limited sample thereby
creating a straightforward sample to analyze.

The next section of the paper briefly discusses the CNOC2 sample and
the volume limited subsample of high-luminosity galaxies.  Section 3
describes in detail our estimate of the projected correlation function
and its errors, along with its sensitivity to various possible
systematic errors. Section 4 reports our best estimates of the
evolution of high-luminosity galaxy clustering in a variety of
cosmologies.  The results are discussed and conclusions drawn in
Section 5. Throughout this paper we use $H_0=100h \kmsm$ as various
cosmologies as specified.

\section{The CNOC2 Sample}

The Canadian Network for Observational Cosmology (CNOC) field galaxy
redshift survey is designed to investigate nonlinear clustering
dynamics and its relation to galaxy evolution on scales smaller than
approximately 20\hmpc\ over the $0\le z \le 0.7$ range.  There is
substantial galaxy evolution over this redshift range
\citep{bes,ldss,cfrs_lf,cowie,huan_lf} 
for which the physical cause is unclear.  The issues in designing a
dynamically useful redshift survey are centered around efficiently and
economically obtaining sufficient data to conclusively answer the
questions posed about the evolution of clustering in galaxies and the
dark matter.  At these relatively modest redshifts the galaxies have
spectra whose lines are within the range where efficient multi-object
spectrographs allow velocities with a precision comparable to local
surveys can be obtained.  The observational procedures build on those
for the CNOC1 cluster redshift survey \citep{yec}, although there are
considerable differences of detail.  The strategy and procedures are
discussed in greater detail in the CNOC2 methods paper
\citep{yee2}.

A representative volume in the universe must contain a reasonable
number of low richness clusters and a good sampling of 50
\hmpc\ voids, or equivalently the ``cosmic'' variance from one sample
to another of equivalent design, but different sky positions, should
not be too large. Taking a CDM spectrum (with $\sigma_8=1$ and
$\Gamma_s=0.2$; see Efstathiou, Bond \& White 1992) as a guide we find
that even a $50
\hmpc$ sphere, with enclosed volume of about $0.5 \times 10^6 h^{-3}\,
{\rm Mpc}^3$, has an expected dispersion in galaxy numbers of
approximately 17\%.  However, a spherical geometry is an ideal not
available to relatively narrow angle surveys. The best available
option is to spread the area of the survey over several independent
patches on the sky. Each patch should subtend an angle exceeding the
correlation length. At $z=0.4$ a patch 0.5 degree across subtends 8.8
\hmpc\ (co-moving, for $\Omega_M=0.2, \Omega_\Lambda=0$) which turns out to be about two
correlation lengths. We can crudely approximate the survey as a set of
cylinders or spheres arranged in a row. A sphere of radius the
correlation length (whatever that happens to be) has an expected
variance of about $3/(3-\gamma)$ in the numbers of galaxies
\citep{lss}, where $\gamma=1.8$ is the approximate slope of the power
law portion of the correlation function.  Over our redshift range
there will be about 200 of these spheres in our sample, which if we
divide into bins of 50, statistically reduces the variance in binned
counts to about 30\%.  The observational practicalities of always
having a field accessible at modest airmass, control over patch to
patch variations, and the constraint of not unduly fragmenting the
survey area, indicate that we need a minimum of four and no more than
about eight of these patches. We choose four.  The resulting survey is
well suited to measuring the evolution of correlations.

The CNOC2 survey is contained in four patches on the sky. Each patch
consists of a central block, roughly 30 arcminutes on a side, with two
``legs'', 10 arcminutes wide and about 40 arcminutes in extent, to
provide an estimate of the effects of structure on larger scales. The
resulting total sky area is about 1.55 square degrees.  The sampled
volume is about $0.5\times 10^6 h^{-3} {\rm Mpc}^3$, roughly
comparable to the low redshift CfA survey used for similar
measurements at low redshift
\citep{dp} which had 1230 galaxies in the ``semi-volume limited''
Northern sample from which the correlation length was derived.  The
redshift range targeted, 0.1 to about 0.7, suggests we set the
limiting magnitude at $R=21.5$ magnitude, which gives a median
redshift of about 0.4. At this limit the sky density is about 6000
galaxies per square degree yielding a photometric sample of about
10,000 galaxies to the spectroscopic limit.

Photometry is obtained in the UBVRI bands, with the R-band fixing the
sample limit at 21.5 mag. The R filter has the important feature of
always being redward of the 4000\AA\ break over our redshift range.
The other bands provide information useful for determining appropriate
k-corrections and separating galaxies into types of different
evolutionary state (an issue not considered in this paper). The
spectra are band-limited with a filter that restricts the range to
4390-6293\AA. The S/N and spectral resolution give an observed frame
velocity error of about 100 $\kms$, as determined by comparison of
independent spectra of the same objects. In total there are about 6000
galaxies with redshifts in our sample.

At low redshift we will use the Las Campanas Redshift Survey (LCRS) to
provide a directly comparable sample. The LCRS is an R-band selected
survey \citep{lcrs} that covers the redshift range 0.033 to 0.15, with
R magnitudes restricted to 15.0 and 17.7 mag. The bright magnitude
limit leads to higher luminosity galaxies being depleted at low
redshifts.  The LCRS's selection function varies from field to field
and under samples galaxies with separations less than about 100 \hkpc.
We compute both magnitude and geometric weights using the approach
adopted in the CNOC cluster redshift survey \cite{yec} but smoothing
over a circle of 0.4 degree. The same procedures are used for the
CNOC2 data, using a smoothing radius of 2 arcminutes. The same
correlation analysis programs were used for the LCRS and CNOC2
data. The only differences are that the LCRS data are not k-corrected
and are analyzed for a single cosmological model, $\Omega_M=0.2, \Omega_\Lambda=0$.

\subsection{The High-Luminosity Subsample}

In order to determine how clustering evolves we must measure the
correlations of a statistically identical population of galaxies at
increasing redshifts, which demands a secure statistical knowledge of
the evolution of the galaxy luminosity function. Individual galaxy
luminosities change through stellar evolution, new star formation and
merging with other galaxies. All of this would have no effect if
galaxy correlations were independent of their mass and luminosity.
However, there is a strong theoretical expectation that galaxy
clustering will increase with mass and hence luminosity
\citep{kaiser84} and there is growing evidence that the effect is
observationally present at both low \citep{loveday} and intermediate
redshifts \citep{roysoc}.  The practical issue is to define samples at
different redshifts which can be sensibly compared. If the evolution
is purely in luminosity, then we would want to compensate for the
luminosity evolution so that the sample limit brings in galaxies of
the same intrinsic luminosity at all redshifts. This would identify
the same galaxies at all times.  If evolution was pure merging, with
no star formation, then it is particular interest to compare the
clustering of galaxies of the same total stellar mass at different
redshifts.  As a currently practical stand-in for stellar mass we use
the k-corrected and evolution compensated R-band absolute luminosity.

Our study of the CNOC2 luminosity function evolution \citep{huan_lf}
found that the R-band evolution of the k-corrected galaxy luminosity
function could be approximated as pure luminosity evolution at a rate,
\begin{equation}
M_R^k(z) = M_R^{k,e}-Qz, 
\label{eq:Q}
\end{equation}
with a mean $Q\simeq 1$ \citep{huan_lf}.  Equation~\ref{eq:Q} defines
the k-corrected and evolution-compensated R absolute magnitudes that
we use to select the sample for analysis.  Figure~\ref{fig:lza} plots
$M_R^{k,e}$ as a function of redshift for the entire flux limited,
$m_R\le 21.5$ mag, CNOC2 sample. It is not necessary to know the
fractional completeness for a correlation measurement. Our correlation
analysis will use galaxies with $M_R^{k,e} \le -20$ mag, which defines
a volume limited sample over the $0.1\le z \le 0.65$ range. Our
resulting subsample (for $\Omega_M=0.2, \Omega_\Lambda=0$) contains 2285 galaxies.  The
alternate cosmological models considered below lead to slightly
different absolute magnitudes and sample sizes.

Beyond $z\simeq 0.55$ the limited CNOC2 spectral band pass leads to a
lower probability that a redshift will be obtained for redder galaxies
\citep{huan_lf,yee2}. This may lead to an erroneously low correlation
in this redshift range. However, as far as we can tell from the
correlation statistics, the high-luminosity galaxies for which we do
have redshifts in this range have correlations statistically
consistent with a smooth continuation of those at lower redshift.

The LCRS data are evolution-compensated with the same $Q$ as the CNOC2
data, although at a mean redshift of about 0.1, this makes very little
difference. The resulting low-redshift subsample derived from LCRS
contains 12467 galaxies for the correlation analysis.

\section{Real Space Correlations}

The goal of this paper is to estimate the evolution of the clustering
of a well defined population of galaxies. The CNOC2 sample is designed
to measure nonlinear clustering on scales of 10\hmpc\ and less, where
clustering is quite naturally measured in terms of the two-point
correlation function, $\xi(r)$. This function measures the galaxy
density excess above the mean background density, $n_0$, at distance
$r$ from a galaxy, $n(r) = n_0[1+\xi(r)]$ \citep{lss}.  Measurement of
the real space correlation function $\xi(r)$ is not straightforward
with redshift space data.  The projected real space correlation
function, $w_p$, removes the peculiar velocities of redshift space at
the cost of making a choice for the length of the redshift column over
which the integration is done.  The correlation function is a measure
of the variation in galaxy numbers from one volume to another. The
measurement technique can easily and fairly subtly either artificially
increase or decrease the variation around the estimated mean. Our
survey was designed to have enough redundancy to explicitly test for a
number of these effects.

We will represent the evolving correlations with a double power law
model. That is, over the range of scales that we are investigating the
two point correlation is well represented as a power law $\xi(r|z) =
[r_0(z)/r]^\gamma$.  Furthermore, the evolution of $\xi$ is accurately
described, over the redshift range we investigate, with the
``$\epsilon$'' model, $\xi(r|z) =\xi(r|0) (1+z)^{-(3+\epsilon)}$, if
lengths are measured in physical co-ordinates \citep{gp,ks}. In this
equation the factor of $(1+z)^{-3}$ allows for the change in the mean
density of galaxies due to expansion. Consequently, if the universe
consists of a set of physically invariant clusters in a smooth
background, then $\epsilon=0$. A clustering pattern that is fixed in
co-moving co-ordinates would have $\epsilon=\gamma-3$, approximately
$-1.2$ for our mean $\gamma$.  A positive $\epsilon$ indicates a
decline of the physical clustered density with increasing redshift, as
might be expected if clustering is growing.  Combining these two power
law equations gives the double power law $\epsilon$ model,
$\xi(r|z)=(r_{00}/r)^\gamma (1+z)^{-(3+\epsilon)}$, in physical
co-ordinates, or, rewritten in co-moving co-ordinates,
\begin{equation}
r_0(z) = r_{00} (1+z)^{-(3+\epsilon-\gamma)/\gamma}.
\label{eq:emodel}
\end{equation}

The following sections discuss the steps leading up to our measurement
and modeling of the co-moving $r_0(z)$.  First we measure the
unclustered distribution $n(z)$, then the projected real space
correlation function $w_p(r_p)$ (after choosing the appropriate
redshift space integration limit $R_p$), fit those values to the
projection of $\xi(r)=(r_0/r)^\gamma$, estimate random errors, do a
$\chi^2$ fit to determine the evolution parameter $\epsilon$ and
finally discuss the systematic errors.

\subsection{The Unclustered Distribution}

A crucial operational detail of correlation measurements is to
accurately assess the mean unclustered density as a function of
redshift. Once the smooth $n(z)$ is known, then we follow the usual
procedure and generate a random sample which follows the redshift
distribution of the data as if they were unclustered. We generate
uniform random positions in the sky area occupied by the galaxies, as
approximated by a series of rectangles.  Correlations are then readily
measured as the ratio of the number of galaxy-galaxy pairs to
galaxy-random pairs. The use of a sample uniformly distributed on the
sky assumes that there are no angular selection effects.  For
instance, crowding of spectrographic fibres causes the LCRS to be
strongly under-sampled for pairs closer than 100\hkpc. The CNOC2
spectroscopic sample was double masked to try to fairly sample all
pair separations.  Slit crowding still leads to some under sampling in
CNOC2 \citep{yee2}.  Consequently we leave these small scales out of
all our fits.  To minimizes other geometric effects, we apply to both
samples an explicit geometric weight, calculated following the
procedures of Yee, Ellingson \& Carlberg (1996). Comparison with
unweighted correlations shows that this makes little practical
difference to the resulting CNOC2 correlations, since the corrections
are 10\% or so in the mean.

One approach to generating an unclustered distribution in redshift is
to use the luminosity function, which can be estimated using
maximum-likelihood techniques that are insensitive to the clustering.
The complication here is that the luminosity functions need to be
generated taking into account the magnitude selection function, in
which only about half of the galaxies within the photometric limit
have redshifts. Moreover, the selection function is magnitude
dependent.  A more direct approach is to model directly the redshift
distribution of the selected subsample. We chose a model function
having sufficiently few parameters that it is not very sensitive to
the details of the clustering that are present. As our
fitting function we adopt a Maxwellian form,
\begin{equation}
n(z|\sigma_z,z_p) = n_0 z^2 
	\exp{\left[-{1\over2}\left({{z-z_p}\over \sigma_z}\right)^2\right]},
\label{eq:nz}
\end{equation}
where $\sigma_z$ and $z_p$ are fitting parameters.  This function was
arrived at after trying various combinations of exponential cutoffs
and power law rises at low redshifts, including log-normal types of
distributions. The resulting form is both simple and adequately
describes the data.  We use a maximum likelihood approach to find the
parameters of this function. The logarithm of the likelihood is
$\log(L) = \sum
\log(L_i)$, where the individual likelihoods are,
\begin{equation}
L_i(z|\sigma_z,z_p) ={ {n(z|\sigma_z,z_p)}\over {\int_{z_b}^{z_t}
n(z|\sigma_z,z_p) \,dz}} .
\label{eq:like}
\end{equation}
The redshift range $z_b$ to $z_t$ for the fit is taken as 0.05 to 0.70
for the CNOC2 sample, although we only use the data between redshift
0.10 and 0.65. The redshift limits are 0.033 and 0.15 for the LCRS.
The CNOC2 redshift distribution in bins of $\Delta z= 0.01$ for all
four patches and the resulting fit is shown in Figure~\ref{fig:cnz} in
bins of $\Delta z= 0.01$.  The best overall fit has $\sigma_z\simeq
0.18$ and $z_p\simeq 0.230$. If the strong clustering feature at
$z\simeq 0.15$ in the 2148-05 patch is not included, then the fits to
the individual fields are all consistent with the global fit.

Inserted into the Figure~\ref{fig:cnz} is a panel showing the 68, 90
and 99\% confidence contours from the maximum likelihood fit. The
error is about 10\% in the $z_p$ parameter and 5\% in $\sigma_z$.
Given these well defined parameters we find that the 68\% confidence
$n(z)$ range is about 10\% at any redshift we use in the analysis.
This is sufficient to measure correlations to 20\% precision with
respect to the background out to about 2 correlation lengths. A small
systematic effect visible in a $\Delta z=0.001$ version of the $n(z)$
plot is that there are small redshift ``notches'' at 0.496 and 0.581
when the [OII] line falls on the 5577\AA\ or 5892\AA\ night sky line,
respectively.  This leads to an underestimate of the true mean density
in the 0.45-0.55 and 0.55-0.65 redshift bins.  This will bias the
derived $r_0$ upwards in these two redshift bins about 2\% and 5\%
respectively, which is within our random errors.

\subsection{The Projected Real Space Correlation Function}

The correlation function is a real space quantity, whereas the
redshift space separation of two galaxies depends on their peculiar
velocities as well as the physical separation.  Although the peculiar
velocities contain much useful information about clustering dynamics,
they are an unwanted complication for the study of configuration space
correlations. The peculiar velocities are eliminated by integrating
over the redshift direction to give the projected correlation
function,
\begin{equation}
w_p(r_p) = \int_{-R_p}^{R_p} \xi(\sqrt{r_p^2+r_z^2})\,dr_z
\end{equation}
\citep{dp}. If we take
a power law correlation $\xi(r)=(r_0/r)^\gamma$ and
integrate to $R_p=\infty$ we find,
$w_p(r_p)/r_p=
\Gamma(1/2)\Gamma((\gamma-1)/2)/\Gamma(\gamma/2)(r_0/r_p)^{\gamma}$
\citep{lss}. However, in a practical survey, summing over ever
increasing distances leads to little increase in the signal and
growing noise from fluctuations in the field density.  The
signal-to-noise considerations in the choice of $R_p$ are
straightforward. To capture the bulk of the correlation signal, $R_p$
should be significantly larger than the local $r_0$ and the length
corresponding to the pairwise velocity dispersion,
$\sigma_{12}/H(z)$. These are both about 3 or 4 \hmpc.  Large values,
say $R_p\simeq 100 \hmpc$, might more completely integrate the
correlation signal but they do so at the considerable cost of
increased noise. Exactly where to terminate the integration depends
greatly on the range of correlations of interest. Here we are focussed
on the non-linear correlations, $\xi>1$.  Before we evaluate an
appropriate choice for $R_p$ we must choose a correlation function
estimator.

\subsection{Galaxy-Galaxy Clustering}

The optimal choice of a statistical estimator of the correlation
function depends on the application.  With point data the basic
procedure is to determine the average number of neighboring galaxies
within some projected radius, $r_p$, and redshift distance $R_p$. The
$ij$ pair is weighted as $w_i w_j$ and the sum over all sample pairs
is $DD$ \citep{lss}.  A random sample of redshifts following the
fitted $n(z)$ is generated along with $xy$ co-ordinates in the visible
sky area of the catalogue.  We then compute the average number of
random sample galaxies within precisely the same volume, assigning the
random points unit weight. This average is known as DR.  Then, we
estimate $w_p(r_p)$ using the simplest and computationally inexpensive
estimator,
\begin{equation}
w_p = { DD\over DR} -1,
\label{eq:dr}
\end{equation}
which is accurate for the nonlinear clustering examined here and
faster than methods which include the RR, \ie\ random-random pairs..
We have verified that the $DD/RR-1$ and $(DD-2DR+RR)/RR$ estimators
give virtually identical results over the range of pairwise
separations that we use in the fits, $0.16\le r_p \le 5.0\hmpc$. These
alternate estimators are known to be superior when $\xi\le 1$, which
only occurs at the outer separation limit of our measurements. We use
100,000 random objects per patch, distributed over the redshift range
0.10 to 0.65 using the fitted $n(z)$.

The DD and DR sums extend over all four patches, so that patch to
patch variations in the mean volume density become part of the
correlation signal. This procedure assumes that there are no
significant patch-to-patch variations in the mean photometric
selection function, which is supported by the absence of any
significant differences in the number-magnitude relation from patch to
patch. We use geometric weights alone for the results presented
here. Magnitude weights give statistically identical results for the
same sample of galaxies, but, as might be anticipated, the errors in the
resulting determination of the correlation evolution are nearly a
factor of two larger.

Estimated projected correlation functions, in co-moving co-ordinates
using $R_p=10\hmpc$, are displayed for the LCRS galaxies bounded by
redshifts [0.033, 0.15] and seven somewhat arbitrary redshift bins for
the CNOC2 data, [0.10, 0.20, 0.26, 0.35, 0.40, 0.45, 0.55, 0.65] in
Figure~\ref{fig:xi_2}.  These boundaries make the end bins bigger than
the middle ones to reduce the variation in numbers between the
bins. For the open model we found the summed geometric weights in the
bins to be [151.8, 264.4, 267.3, 471.6, 301.2, 496.7, 286.2], and
[172.2, 305.0, 321.1, 595.6, 368.3, 600.3, 315.4] for the $\Lambda$
cosmology, showing the sample differences due to cosmology are quite
small.  Extensive testing found that provided the bins are not made
significantly narrower than the adopted limits, the bin sizes make no
significant difference to the results.  Adjusting the bins to have
nearly constant numbers makes no difference to our result. Of course
the LCRS data is very important for providing a solid measurement at
low redshift.

The measured $w_p$ are fit to the projection of the power law
correlation function, $\xi(r)=(\hat{r}_0/r)^{\hat{\gamma}}$, estimating
both $\hat{r}$ and $\hat{\gamma}$. The errors at each $r_p$ are taken
as $1/\sqrt{DD}$.  The fits are restricted to the $0.16\le r_p
\le 5.0\hmpc$ range where there are minimal complications from geometric
selection effects and the correlation signal is
strong. Figure~\ref{fig:xi_2} displays these fits as solid lines. Also
shown in Figure~\ref{fig:xi_2} as dashed lines are fits where we have
converted to standardized $\gamma=1.8$ correlation lengths, $r_0$,
using $r_0 = \hat{r}_0^{\hat{\gamma}/1.8}$.  All results here are
derived using co-moving co-ordinates, and normalized to a Hubble
constant $H_0=$ $100 h \kmsm$. The results displayed in
Figure~\ref{fig:xi_2} are derived assuming a background cosmology of
$\Omega_M=0.2,
\Omega_\Lambda=0$.

\subsection{Random Errors of the Correlations}

The problem of error estimates for correlation measurements remains a
topic of active research. The shot noise estimate of the
fractional error as $1/\sqrt{DD}$ is appropriate for weak clustering,
but a substantial underestimate for strongly nonlinear clustering,
where the clustering itself reduces the effective number of
independent pairs. A formal error expression in terms of the three-
and four-point correlation function is available \citep{lss} but
cumbersome and computationally expensive. Resampling techniques, such
as the Bootstrap and Jackknife \citep{et}, produce substantial
over-estimates of the error.

A straightforward approach to error estimates is to take advantage of
our sample being distributed over a number of separate patches.  We
separately fit each of the four CNOC2 and six LCRS patches, to obtain
an $r_0$ for each patch or slice. The estimated error in any
correlation length, $r_0$, is simply,
\begin{equation}
\sigma^2_{r0}= {1\over {n-1}}\sum_{i=1}^n 
	[r_0(i)-\langle r_0 \rangle]^2,
\label{eq:err}
\end{equation}
where the sum extends over the $n=4$ CNOC2 patches and $n=6$ LCRS
slices at the redshift of interest. The average correlation length in
Eq.~\ref{eq:err} is computed from the individual patches and is not
equal to the correlation of the four fields combined, which is
generally larger than the average since it includes the patch to patch
variation in mean counts as part of the signal. Because we have only
four CNOC2 patches and six LCRS strips the estimated errors will
themselves have substantial fluctuations. The resulting co-moving
correlation lengths for a power law model are displayed for a range of
$R_p$ in Figure~\ref{fig:xiR}. The open circles are the results for
the four individual CNOC2 patches (the individual LCRS slices are so
similar that they are not displayed).  The solid points give the
result from the combined data, along with the estimated error. The
mean of the fitted slopes is $\gamma=1.87\pm0.07$.  In principle our
smaller values of $R_p$ could cause the measured $w_p$ to miss real
correlation at large $r_p$, leading to $\gamma$ values that are systematically
to large. However, for the $R_p \ge 10\hmpc$ our fitted $\gamma$ values
have no significant dependence on $R_p$.

\subsection{Observational Estimates of $\epsilon$}

The co-moving $r_0(z_j)$, derived from fits to the measured $w_p$, are
in turn fit to the $\epsilon$ model by minimizing,
\begin{equation}
\chi^2 = \sum_j \left[
	{{r_0^{\gamma}(z_j)(1+z_j)^{3-\gamma} 
	- r^\gamma_{00} (1+z_j)^{-\epsilon}}
	\over {\sigma_{\xi}(z_j)}}\right]^2,
\end{equation}
over the $j$ redshift bins by varying $r_{00}$ and $\epsilon$. The
quantity $r_0^{\gamma}(z_j)(1+z_j)^{3-\gamma}$ is proportional to the
mean clustered physical density. The $r_0(z_j)$ are the results of the
fits to the correlation measurement of the four patches combined.  The
$\sigma_{\xi}(z_j)$ are the variances estimated from the standard
deviations of the $r_0$ values $\sigma_{r0}$, re-expressed as a
variance of the correlation amplitude, $\sigma_\xi(z) \simeq
[(r_0(z)+\sigma_{r0}(z))^\gamma- r_0^\gamma(z)](1+z)^{3-\gamma}$. The
$\sigma_{r0}$ are evaluated using Equation~\ref{eq:err}. The $\chi^2$
statistic allows us to evaluate absolute goodness of fit, as well as
determine parameter confidence intervals.

\subsection{Systematic Errors of the Correlations}

We can assess the effect of varying $R_p$ in the $w_p$ integration
using the results of the fits to the $\epsilon$ model,
Eq.~\ref{eq:emodel}. Fitted $r_{00}$ and $\epsilon$ are displayed as a
function of $R_p$ in Figures~\ref{fig:rR} and
\ref{fig:eR}. The errors are the 90\% confidence intervals. 
No 90\% confidence fits were found at $R_p$ of 20 and 100 \hmpc, which
likely reflects variations in the estimated errors more than a true
failure of the model. From these two figures we conclude that $R_p=$
10 or 30 \hmpc\ converge to give statistically identical values of
$r_{00}$ and $\epsilon$.  Smaller $R_p$ values fail to include the
full signal and larger $R_p$ values give huge patch to patch
variations as large voids come and go. The most conservative choice
for $R_p$ is 10\hmpc, the one with the largest error in the stable
range. Figure~\ref{fig:rR} weakly suggests that somewhat larger $R_p$
would lead to a small increase in the correlation length, to
$r_{00}\simeq 5.2\hmpc$. We will adopt the $R_p=10\hmpc$ fits as our
standard results, noting that the inferred $\epsilon$ have essentially
no dependence on $R_p$.

For small survey volumes the derived correlation length tends to
systematically underestimate the result from a very large area. That
is, clustering is known to be significant on scales of at least
100\hmpc, hence surveys smaller than that in any dimension are likely
to be measuring the range of clustering about either a local valley or
plateau, and not seeing the full range of clustered density.  The
effect of increasing survey size on correlations can be seen in
Figure~\ref{fig:cr0z}, in that the combined analysis (filled circles)
generally gives correlations higher than the mean of the individual
patches (other symbols).  Quantitatively, the straight mean of the
CNOC2 $r_0$ is 3.2\hmpc; the median is 3.4\hmpc. It is more
appropriate to average together the pair counts, which is equivalent
to taking the average $\langle r_0^\gamma\rangle^{1/\gamma}$ leading
to an average correlation length of 3.5\hmpc. Performing a joint
correlation analysis of all four patches together gives an $r_0$ of
4.0\hmpc.  This raises the question as to whether the correlations
have converged within the current survey. The expected variation from
patch to patch for the given volumes with narrow redshift bins is
about 45\%, which is consistent with the difference between a
correlation length of 3.5 and 4.3\hmpc. In the combined sample with
larger bins we expect that there could be as much as about 10\% of the
variance missing, which would boost the correlation lengths by another
5\%.

\section{The Evolution of Galaxy Clustering}

The correlation lengths for CNOC2 and LCRS, derived from
fitting $w_p(r_p)$ as discussed in \S3.3 and analyzed in precisely the
same way for our standard $R_p=10\hmpc$ and $Q=1$, are shown in
Figure~\ref{fig:cr0z} and reported in Table~\ref{tab:r0z}. It is
immediately clear that there is relatively little correlation
evolution for high-luminosity galaxies. It must be borne in mind that
the sample is defined to be a similar set of galaxies with $L\gtrsim
L_\ast$, with luminosity evolution-compensated, that approximates a
sample of fixed stellar mass with redshift. Samples which admit lower
luminosity galaxies, or do not correct for evolution, or are selected
in bluer pass-bands where evolutionary effects are larger and less
certainly corrected, will all tend to have lower correlation
amplitudes.

The $\chi^2$ contours of the $r_{00}-\epsilon$ model fits to the
measured correlations of \S3 are shown in Figure~\ref{fig:r00e}.  At
redshifts beyond 0.1 or so, the choice of cosmological model has a
substantial effect on the correlation estimates. Relative to a high
matter density cosmological model, low density and $\Lambda$ models
have larger distances and volumes, which cause the correlations to be
enhanced. The LCRS data are analyzed only within the $\Omega_M=0.2, \Omega_\Lambda=0$
model. The correlations for three cosmologies, flat matter dominated,
open, and low-density $\Lambda$, are shown in
Figure~\ref{fig:r0z}. The $\chi^2$ contours at the 68\%, 90\% and 99\%
contours are shown in Figure~\ref{fig:r00e}. The best fit $\epsilon$
value is $-0.17\pm 0.18$ for $\Omega_M=0.2,\Omega_\Lambda=0$ with
$r_{00}=5.03\pm0.08\hmpc$.  The evolution rates for the flat matter
dominated and flat low-density models are $\epsilon=+0.8\pm0.22$ and
$\epsilon=-0.8\pm0.19$, respectively, with $r_{00}$ of
$5.30\pm0.1\hmpc$ and $4.85\pm0.1\hmpc$, respectively. These are
marked with plus signs in Figure~\ref{fig:r00e}.

The effects of alternate values for the luminosity evolution are shown
in Figure~\ref{fig:r00e} with crosses indicating the results for $Q=0$
and $Q=2$, with the adopted value being $Q=1$. The absolute magnitude
limit remains $M_R=-20$ mag in all cases and we use the $\Omega_M=0.2,
\Omega_\Lambda=0$ cosmology. For $Q=0$ the
summed bin weights are [176.1,329.2,347.7,659.7,417.7,673.6,318.4] and
for $Q=2$ they are [128.7,203.7,190.2,317.6,191.2,296.0,159.3]. The
effect is that less (more) evolution compensation gives rise to a more
(less) rapid decline in the correlations with increasing redshift.  In
the absence of any allowance for luminosity evolution, $Q=0$, galaxies
of lower luminosity are included with increasing numbers at higher
redshift. Galaxy correlations tend to decline slightly with decreasing
luminosity with the strongest declines being at high-luminosity
\citep{loveday} although the details of this important effect remain
controversial. A preliminary investigation finds a small effect in the
CNOC2 sample
\cite{roysoc}. If we do not correct for luminosity evolution, then
intrinsically lower luminosity galaxies are included in increasing
numbers at higher redshifts, which will leads to an increased rate of
decline of correlations with redshift, as we have empirically
demonstrated here.  The observed effect over the 0.0 to 0.65 redshift
range is approximately $\Delta \epsilon \approx -0.3 \Delta Q$. This
effect is partially responsible for the difference between the results
here and those of LeF\`evre \etal\ (1996).

\section{Comparison of Observations and Theory}

We now can compare our measurements of clustering evolution to the
various simple theoretical models and analytic fits to n-body
simulation results. We will cast these predictions into
the form of an equivalent theoretical $\epsilon_T$ using,
\begin{equation}
\xi(r,z_2) = \xi(r,z_1) \left({{1+z_2}\over{1+z_1}}
	\right)^{-(3+\epsilon_T -\gamma)}, 
\label{eq:epst}
\end{equation}
where we have assumed that the correlation function will always be of
the form $r^{-\gamma}$.  Linear growth is the simplest case,
\begin{equation}
\xi(r,z) = D^2(z,\Omega) \left({r_{00}\over r }\right)^\gamma,
\label{eq:lin}
\end{equation}
where $D(z,\Omega)$ is the linear perturbation growth factor
\citep{lss,ppc}. For the 
$\Omega_M=1$, $\Omega_\Lambda=0$, $D(z)$ is simply the expansion
factor, $a(z)=(1+z)^{-1}$, which gives the result that
$\epsilon_T=0.8$. For other $\Omega$ values we approximate the
redshift dependence as a power law in $1+z$ by evaluating the growth
factor at $z=0$ and $z=0.5$, as is appropriate for this survey.  The
results are presented in Table~\ref{tab:eps}.

An alternate clustering model is to allow for some bias, $b(z)$, of
galaxies clustering with respect to the dark matter,
$\xi_{gg}=b^2\xi_{\rho\rho}$.  There are two simple forms which
describe the bias, $b(z)$, \citep{mowhite},
\begin{equation}
b_{MW}(z) = 1 -{1\over \delta_c}+{\delta_c \over{D^2(z)\sigma^2(M)}},
\label{eq:mw}
\end{equation}
where $\sigma(M)$ is the tophat mass variance \citep{bbks} in a sphere
of radius $R=1/\Omega_M^{1/3}\hmpc$ and the critical linear overdensity,
$\delta_c\simeq 1.68$.  A refinement to this formula based on fitting
to n-body results is \citep{jing},
\begin{equation}
b_{J}(z) =  b_{MW}(z)\left(1+{{0.5 D^4(z)\sigma^4(M)}\over \delta_c^4}\right)
	^{0.06-0.02n},
\label{eq:jing}
\end{equation}
where $n=d\ln{\sigma^2(R)}/d\ln{R}-3$ is the effective index of the
perturbation spectrum. We use the fit to CDM spectrum of Efstathiou,
Bond \& White (1992) to evaluate our tophat variances.

One possibility of relevance only in an $\Omega_M=1$ cosmology, is
that clustering obeys a scaling law \citep{lss,EDWF},
$\xi(r,t) = \xi(s)$, with 
\begin{equation}
s \propto r (1+z)^{2/(n+3)}.
\label{eq:scaling}
\end{equation}
CDM has such a large negative effective index, $n\simeq -2.1$ on
galaxy scales, that it gives rise to a very large theoretical value of
$\epsilon_T\simeq 7$, as was seen in the early CDM simulations
\citep{DEFW}.  An $\epsilon_T$ this large is completely excluded by
clustering evolution studies.

The observations and theoretical predictions in Table~\ref{tab:eps}
allow us to draw a number of conclusions.  Linear theory is not a very
good model because of both biasing and nonlinearities, given that the
range of scales fit in the power law spans overdensities ranging from
$10^3$ to $0.2$, but it is a useful reference point.  The open and
$\Omega_M=1$ cosmologies are consistent with linear theory growth but
the low density $\Lambda$ model is marginally excluded at 3.4 standard
deviations (s.d.\,).  A comparison to n-body experiments
\citep{ccc,ckkk,kk99} shows the same approximate consistency with
low density mass-traces-light cosmologies. Mo \& White biasing in an
open cosmology is marginally excluded at 3.8~s.d.\ and Jing biasing
more conclusively at 4.7 s.d.  Biased clustering in an $\Omega_M=1$
cosmology is excluded at more than 6~s.d.  The low density flat model
is acceptable under all biasing models. Bearing in mind that the
evolution of correlations is a test of both a galaxy formation and
evolution model and the cosmology, these results mainly exclude models
where galaxies are closely identified via the biasing mechanism with
dark matter halos in an $\Omega_M=0.2, \Omega_\Lambda=0$, cosmology or
in an $\Omega_M=1$ cosmology. The problem in both cases is that they
predict almost no correlation evolution, whereas we observed a small
but significant decrease of the co-moving correlation length with
redshift.

\section{Conclusions}

The CNOC2 redshift survey has measured precision velocities for more
than 6000 galaxies in the redshift 0.1 to 0.7 range. The sky area of
about 1.55 square degrees therefore covers a volume of about
$0.5\times 10^6 h^{-3} \, {\rm Mpc}^3$. We have defined a volume
limited subsample of those galaxies with k-corrected and evolution
corrected R-band absolute magnitudes of $M_R^{k,e} \le -20$ mag, where
$M_\ast \simeq -20.3$ mag in the R-band. This subset contains about
2300 galaxies in the 0.1 to 0.65 redshift range. At low redshift we
add about 13000 identically selected galaxies from the LCRS.

The correlation measurements from this paper are contained in
Figure~\ref{fig:cr0z} and the associated Table~\ref{tab:r0z}.  Over
the redshift range examined, the correlation evolution can be
described with the double power law model, $\xi(r|z) =
(r_{00}/r)^\gamma (1+z)^{-(3+\epsilon-\gamma)}$, in co-moving
co-ordinates. We measure a $\gamma=1.87\pm0.07$ and set $\gamma=1.8$
for fitting purposes.  Our results for various cosmologies and
evolution corrections are shown in Figure~\ref{fig:r00e}.  The primary
conclusion is that correlations show a weak decline with
redshift. Furthermore, there is no evidence in the current data for a
change in the slope of the correlation function with redshift.

These observations test both the amplitude and redshift evolution of
clustering predictions.  They jointly constrain the cosmology and the
galaxy formation history, and do not provide any strong constraints on
the background cosmology by themselves. The comparison of our
measurements and analytic fits is presented in Table~\ref{tab:eps}.
The correlation amplitude and slope are in quite good agreement with
appropriately selected dark matter halos in a CDM simulation
\citep{virgogal}, however the agreement depends fairly sensitively on
the mass range selected \citep{kcdw}. The rate decline of clustering
with redshift is slow but significant in all cosmologies examined
here. That is, $\epsilon$ is always greater than $-1.2$, the value for
a fixed co-moving clustering length.  The prediction of a slower
clustering evolution in a biased $\Omega_M=1$ cosmology is completely
excluded by the more rapid decline measured here and somewhat
marginally excluded in an $\Omega_M=0.2,
\Omega_\Lambda=0$ cosmology.  The models of biased clustering is an
$\Omega_M=0.2, \Omega_\Lambda=0.8$ cosmology are statistically
consistent with our measurements.

\acknowledgments

This research was supported by NSERC and NRC of Canada.  HL
acknowledges support provided by NASA through Hubble Fellowship grant
\#HF-01110.01-98A awarded by the Space Telescope Science Institute,
which is operated by the Association of Universities for Research in
Astronomy, Inc., for NASA under contract NAS 5-26555.  We thank the
CFHT Corporation for support, and the operators for their efficient
control of the telescope.

\newpage
~
\begin{table}[h]
\caption{Redshift Evolution of Correlations \label{tab:r0z}}

\bigskip
\begin{tabular}{|c||c|c|c||c|c|c|}\hline
 & \multicolumn{5}{c|}{$r_0$ in \hmpc\ for $\gamma=1.8$} \\\hline
$\langle z \rangle$ & \multicolumn{3}{c||}{$\Omega_M=0.2,\Omega_\Lambda=0$}
  & $\Omega_M=1.0,\Omega_\Lambda=0$
  & $\Omega_M=0.2,\Omega_\Lambda=0.8$ \\\hline
  & $Q=0$ & $Q=1$ & $Q=2$ & $Q=1$  & $Q=1$ \\ \hline
  0.10 &    4.75 $\pm$ 0.05   &    4.75 $\pm$ 0.05   &    4.75 $\pm$ 0.05   &    4.75 $\pm$ 0.05   &    4.75 $\pm$ 0.05 \\
  0.16 &    4.52 $\pm$ 0.68   &    4.85 $\pm$ 0.82   &    4.93 $\pm$ 1.08   &    4.67 $\pm$ 0.76   &    4.94 $\pm$ 0.85 \\
  0.24 &    4.01 $\pm$ 0.29   &    4.13 $\pm$ 0.35   &    4.03 $\pm$ 0.42   &    3.65 $\pm$ 0.34   &    4.49 $\pm$ 0.44 \\
  0.31 &    3.92 $\pm$ 0.20   &    4.14 $\pm$ 0.31   &    4.01 $\pm$ 0.77   &    3.72 $\pm$ 0.47   &    4.53 $\pm$ 0.21 \\
  0.38 &    3.94 $\pm$ 0.41   &    3.90 $\pm$ 0.37   &    4.35 $\pm$ 0.34   &    3.66 $\pm$ 0.31   &    4.53 $\pm$ 0.28 \\
  0.42 &    3.44 $\pm$ 0.43   &    3.80 $\pm$ 0.59   &    4.28 $\pm$ 0.73   &    3.58 $\pm$ 0.40   &    4.14 $\pm$ 0.64 \\
  0.49 &    3.71 $\pm$ 0.11   &    4.26 $\pm$ 0.18   &    5.01 $\pm$ 0.45   &    3.77 $\pm$ 0.28   &    4.54 $\pm$ 0.12 \\
  0.59 &    3.56 $\pm$ 0.25   &    3.68 $\pm$ 0.27   &    4.75 $\pm$ 0.49   &    3.13 $\pm$ 0.31   &    4.00 $\pm$ 0.34 \\
\hline
\end{tabular}
\end{table}

\newpage
\begin{table}[ht]
\caption{Comparison of Observed and Predicted $\epsilon$ values \label{tab:eps}}
\bigskip
\begin{tabular}{|c||c|c|c|}\hline
$\epsilon$ source
	& $\Omega_M=0.2,\Omega_\Lambda=0$
	& $\Omega_M=1.0,\Omega_\Lambda=0$
	  & $\Omega_M=0.2,\Omega_\Lambda=0.8$ \\\hline
observation & $-0.17\pm 0.18$ & $+0.8\pm0.22$ & $-0.8\pm0.19$ \\
linear & $-0.35$ & $+0.80$ & $-0.15$ \\
MW biased & $-0.88$ & $-0.66$ & $-0.83$ \\
Jing biased & $-1.02 $ & $-1.00$ & $-1.00$ \\
CDM scaling & --- & $+7.23$ & --- \\
\hline
\end{tabular}
\end{table}

\vfill
\clearpage

\figcaption[lza.ps]{
Absolute magnitudes in the R-band, k-corrected and
evolution-compensated ($Q=1$), versus redshift. A volume limited
sample with $M_R^{k,e} \le -20$ mag is used for all the analysis in
this paper.
\label{fig:lza}}

\figcaption[nzfit.ps]{
Maximum likelihood fit of $n(z) \propto z^2 \exp{(-{1\over
2}[(z-z_p)/\sigma_z]^2)}$ to the observed $n(z)$, where $\sigma_z$ and
$z_p$ are the parameters that are varied.  The best fit has
$\sigma_z\simeq 0.18$ and $z_p\simeq 0.230$. The inset figure shows
the 68, 90 and 99\% confidence contours for the maximum likelihood fit
of these data, without binning, to the model function.
\label{fig:cnz}}

\figcaption[xi_2.ps]{
The measured projected correlations, $w_p$, for $\Omega_M=0.2, \Omega_\Lambda=0$, as a
function of redshift. The power law fits with unconstrained slopes are
plotted as solid lines. The projected correlations derived assuming
$\gamma=1.8$ are shown as dashed lines. The lowest line is for the
LCRS sample. The next seven are for the different redshift intervals
of the CNOC2 survey. The point-to-point scatter gives an indication of
the errors. These correlations are calculated with $R_p= 10\hmpc$
(co-moving).
\label{fig:xi_2}}

\figcaption{
The resultant $r_0(z)$ for a range of $R_p$ in the $w_p(r_p)$
integration. From left to right starting at the top the $R_p$ are 5,
7, 20, 50, 70 and 99 \hmpc.  The individual sky patches 0223, 0920,
1447 and 2148, are shown with with plus, asterisk, circle and cross
symbols, respectively, and the solid symbol is the result of combined
patches analysis. The ``standard'' correlations for the adopted value
of $R_p=10\hmpc$ are shown in Figure~\ref{fig:cr0z}.
\label{fig:xiR}}

\figcaption{
The derived $r_{00}$ as a function of the integration length, $R_p$,
used to define $w_p(r_p)$. The 90\% confidence intervals are shown.
Points without error flags have fits that are outside the 90\%
confidence interval. This most likely arises because the variances
used to calculate $\chi^2$ are estimated from the dispersion of the
four patches, which will sometimes lead to erroneously small variances
and hence large $\chi^2$ values.
\label{fig:rR}}

\figcaption{
The derived $\epsilon$ as a function of the integration length, $R_p$,
used to define $w_p(r_p)$. The 90\% confidence intervals are
shown. Points without error flags have fits that are outside the
90\% confidence interval.
\label{fig:eR}}

\figcaption{
The correlation lengths (normalized to $\gamma=1.8$) as a function of
redshift for $\Omega_M=0.2,\Omega_\Lambda=0$. The filled diamond is
for the LCRS sample. The CNOC2 errors are estimated from the variance
of the four sky patches (shown with plus, asterisk, circle and cross
symbols for the 0223, 0920, 1447 and 2148 patches, respectively) and
the six LCRS slices (not shown since the differences are small). The
filled circles are the correlations from the four fields combined.
Note that these are in general always larger than the mean of the
individual fields, since they include field-to-field variance.
\label{fig:cr0z}}

\figcaption{
The $\chi^2$ confidence levels for fits to the $\epsilon$ model for
the $\Omega_M=0.2,\Omega_\Lambda=0$ model. The contours are for 68\%,
90 and 99\% confidence. The plus signs mark the results for
$\Omega_M=1, \Omega_\Lambda=0$, where $\epsilon\simeq 0.8$ and
$\Omega_M=0.2, \Omega_\Lambda=0.8$ where $\epsilon=-0.8$. The crosses
show the outcome for no evolution correction, $\epsilon=0.3$; and
$Q=2$, evolution, $\epsilon=-0.3$.
\label{fig:r00e}}

\figcaption{
The cosmology dependence of the correlation lengths for $\Omega_M=0.2,
\Omega_\Lambda=0$ (circles), $\Omega_M=1, \Omega_\Lambda=0$ (triangles) 
and $\Omega_M=0.2, \Omega_\Lambda=0.8$ (plus signs).  The darkened
points are from the LCRS with $\Omega_M=0.2, \Omega_\Lambda=0$.
\label{fig:r0z}}

\newcounter{figi}
\newcommand{\nfig}{\addtocounter{figi}{1}\thefigi}

\begin{figure}
\figurenum{\nfig}\includegraphics[width=0.8\hsize]{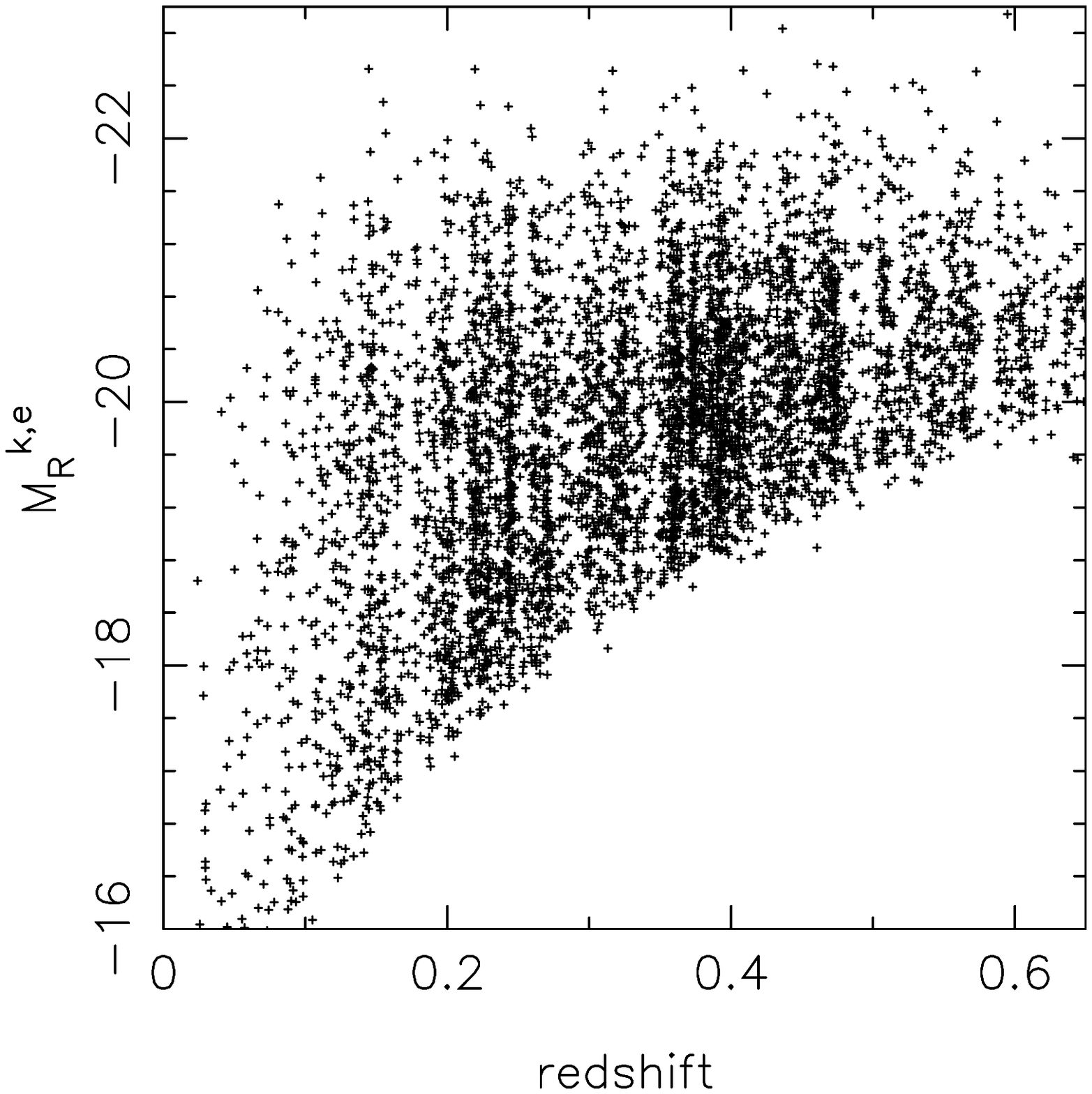} 
\caption{}
\end{figure}  
\clearpage

\begin{figure}
\figurenum{\nfig}
\begin{picture}(0,0)(0,0)
	\put(0,-450){\includegraphics[width=1.0\hsize]{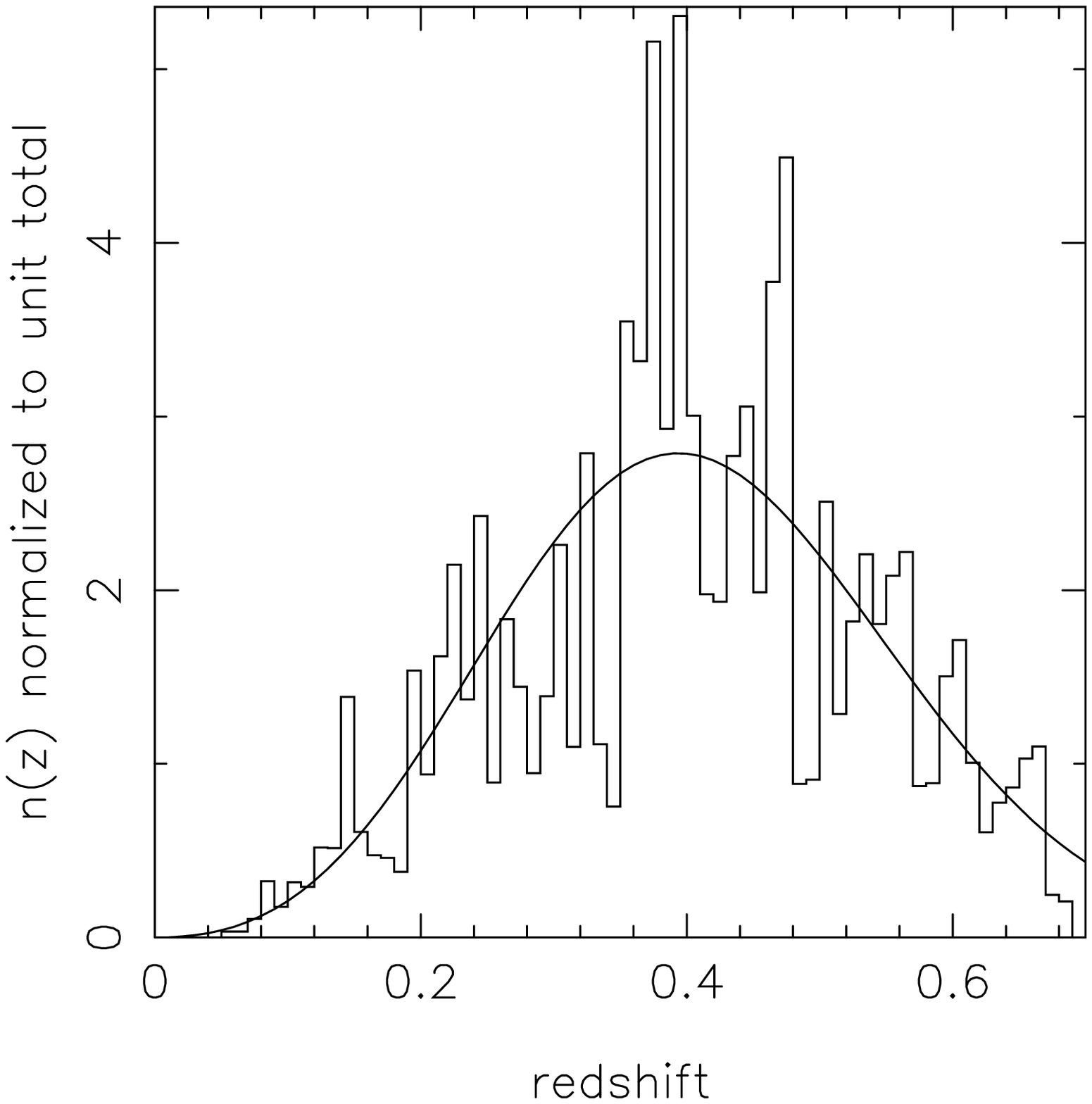} }
	\put(85,-170){\includegraphics[width=0.35\hsize]{f2a.ps}}
\end{picture}
\vskip 16.0truecm
\caption{}
\end{figure}  

\begin{figure}\figurenum{\nfig}
	\includegraphics[height=0.9\vsize]{f3.ps} \caption{}\end{figure}  

\clearpage
~

\begin{figure}
\figurenum{\nfig}
\begin{picture}(0,0)(0,0)
	\put(  0,400){\includegraphics[width=0.4\hsize]{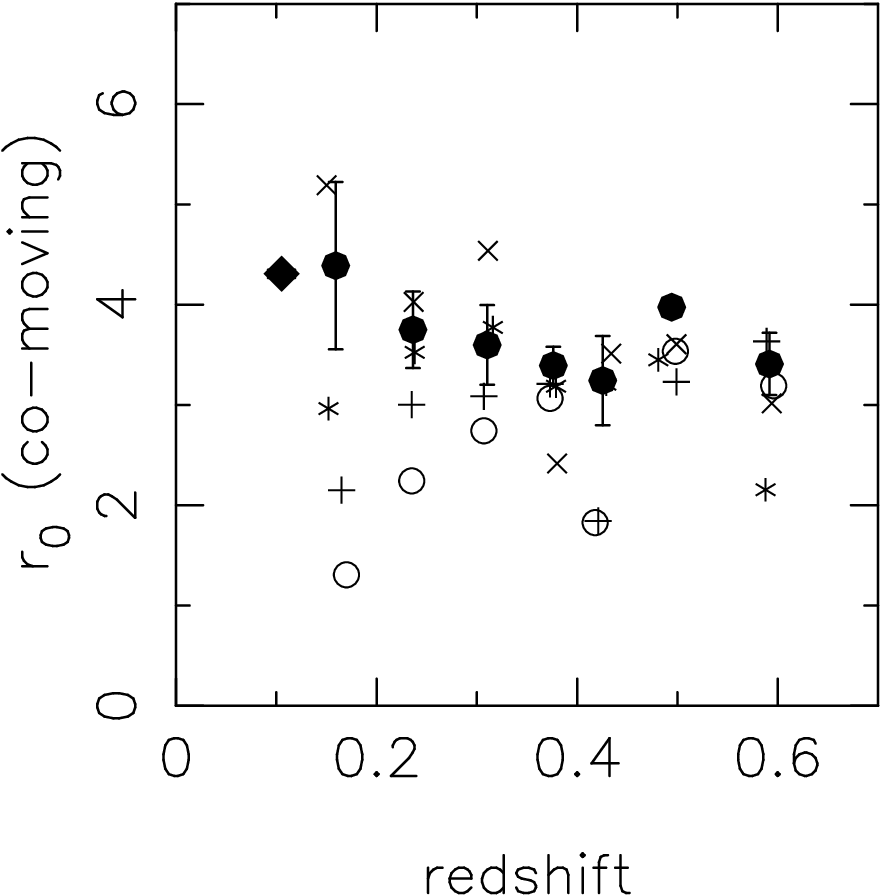} }
	\put(220,400){\includegraphics[width=0.4\hsize]{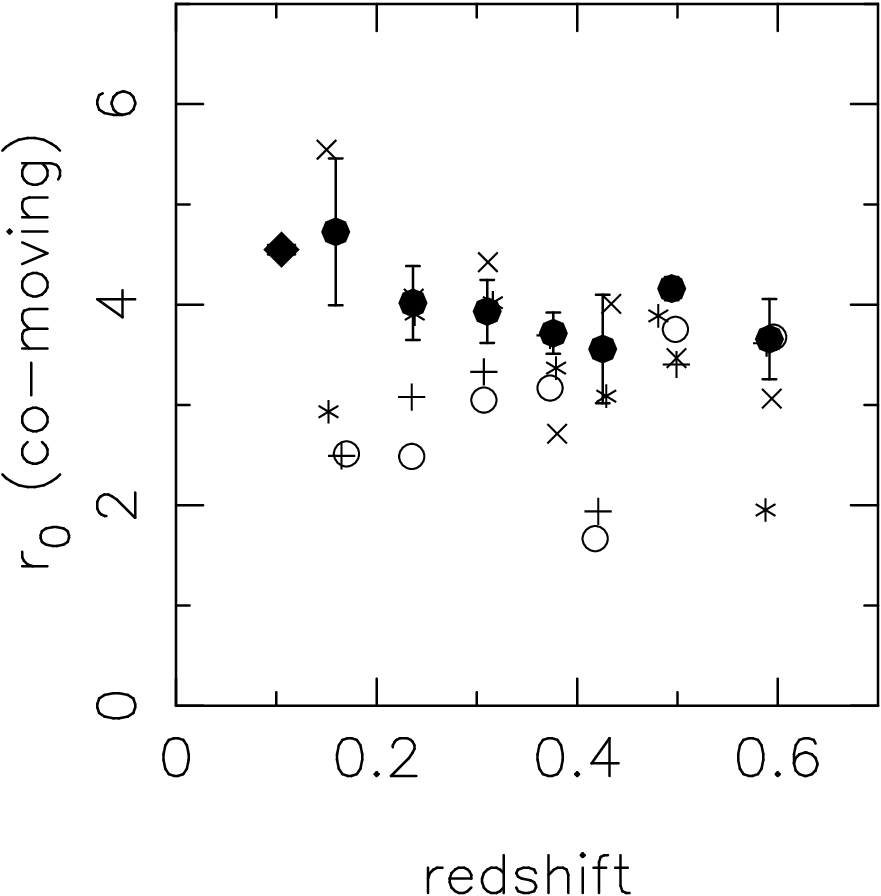} }
	\put(  0,200){\includegraphics[width=0.4\hsize]{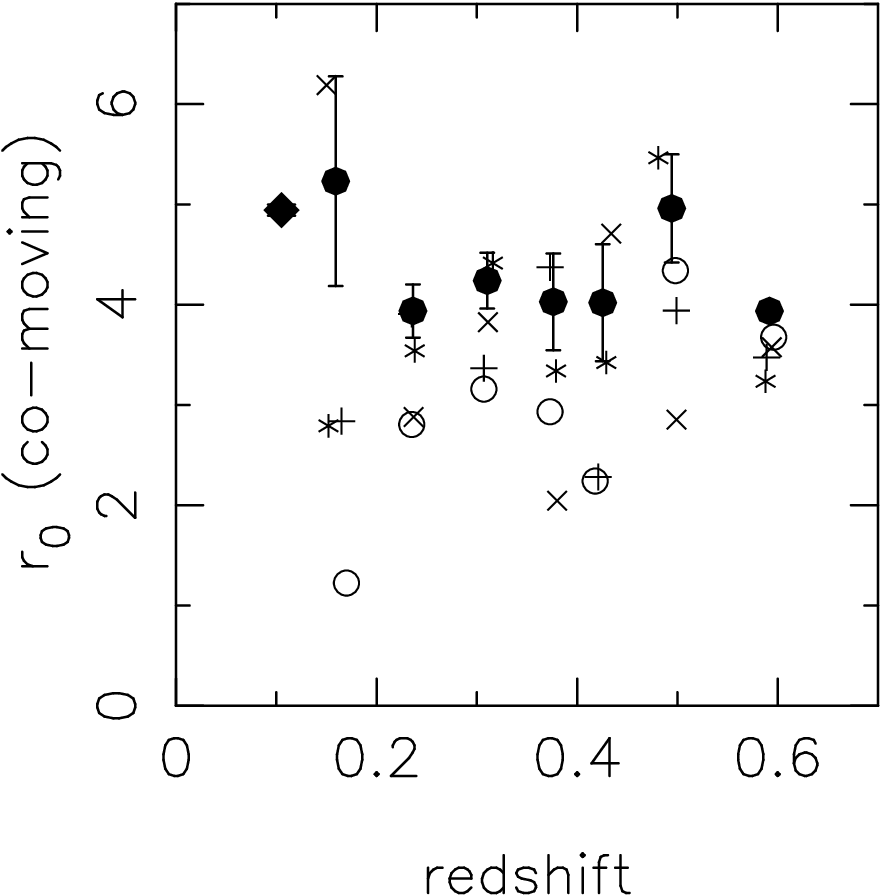} }
	\put(220,200){\includegraphics[width=0.4\hsize]{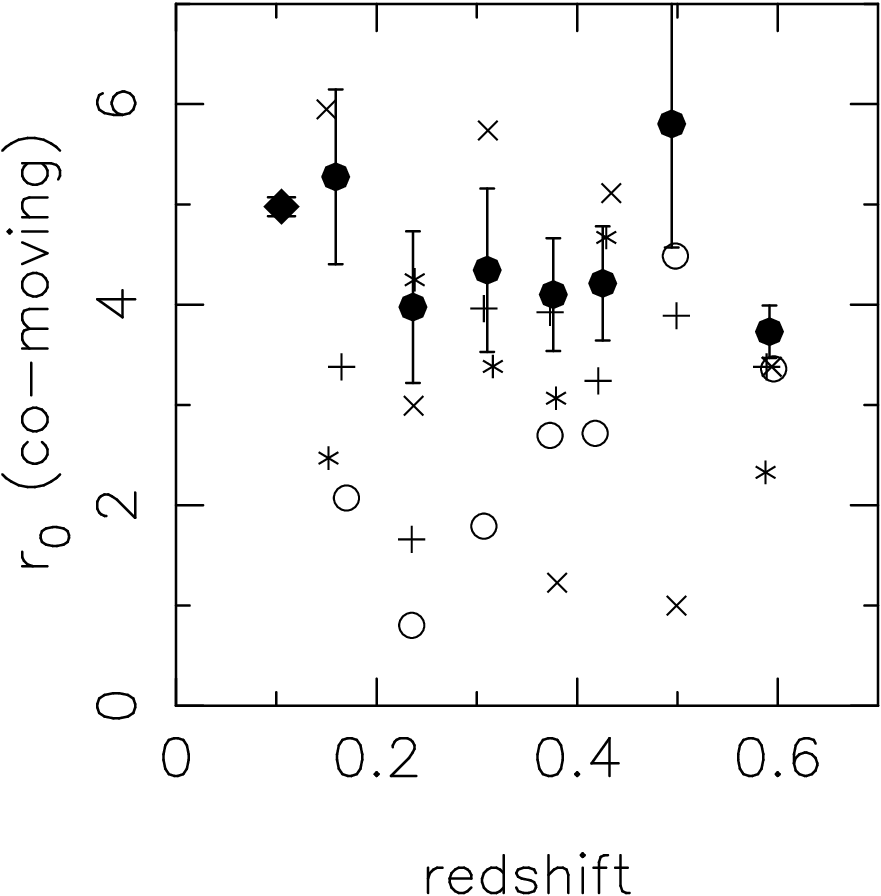} }
	\put(  0,  0){\includegraphics[width=0.4\hsize]{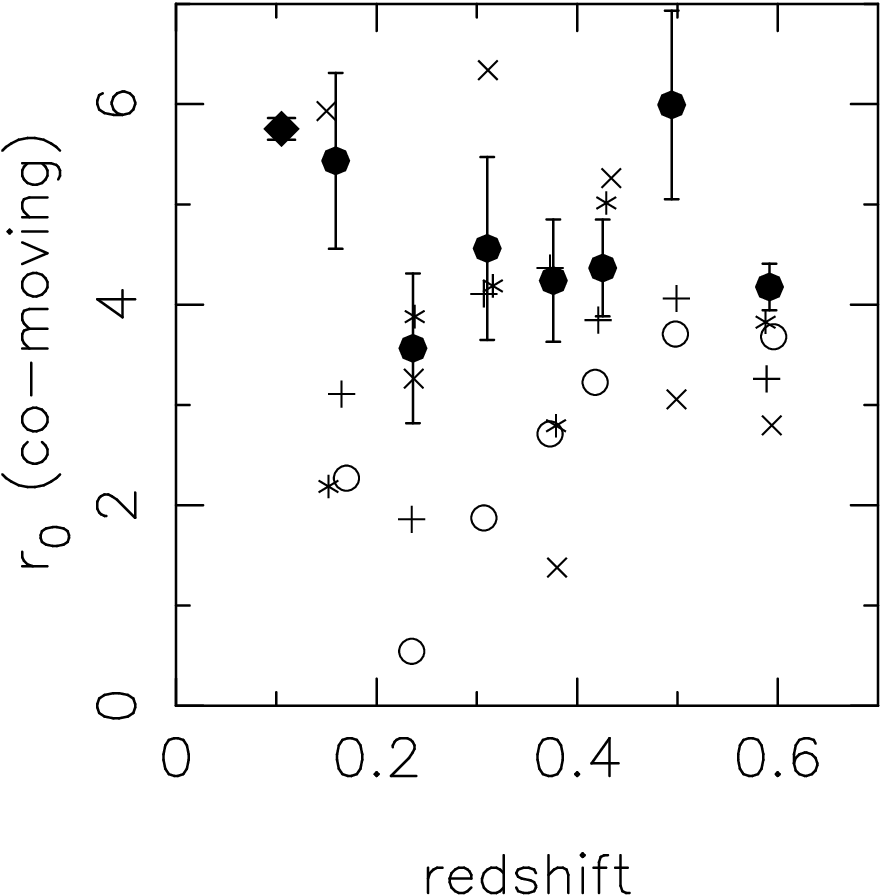} }
	\put(220,  0){\includegraphics[width=0.4\hsize]{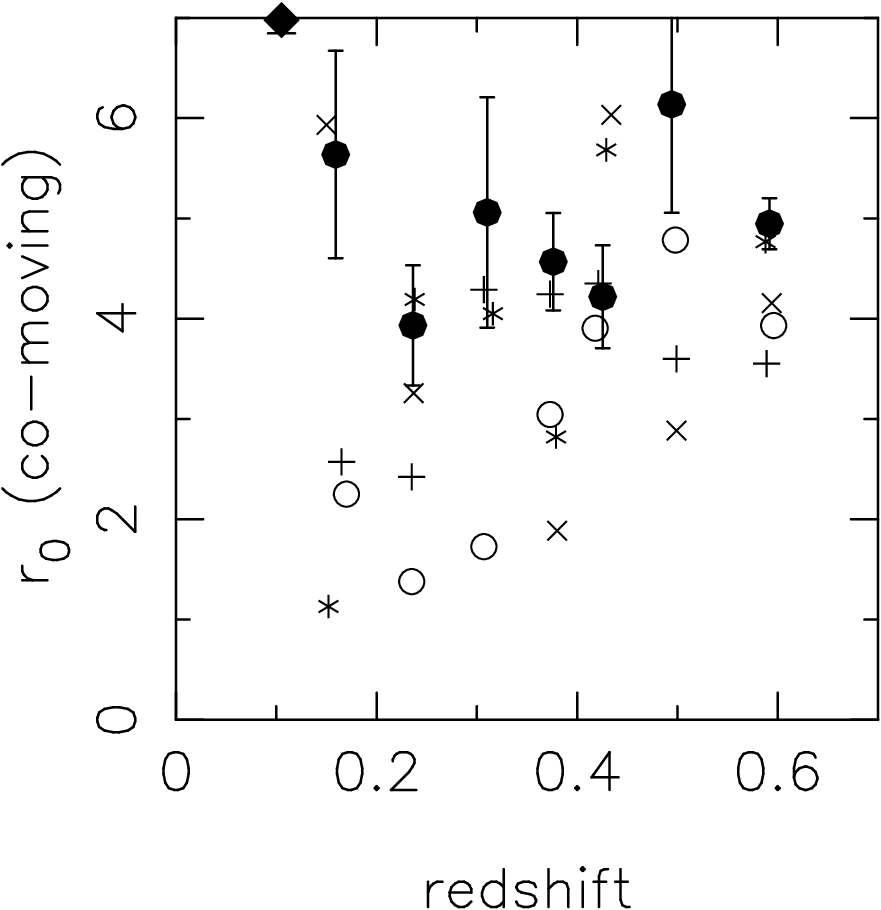} }
\end{picture}
\caption{}\end{figure}  

\clearpage

\begin{figure}\figurenum{\nfig}
	\includegraphics[width=\hsize]{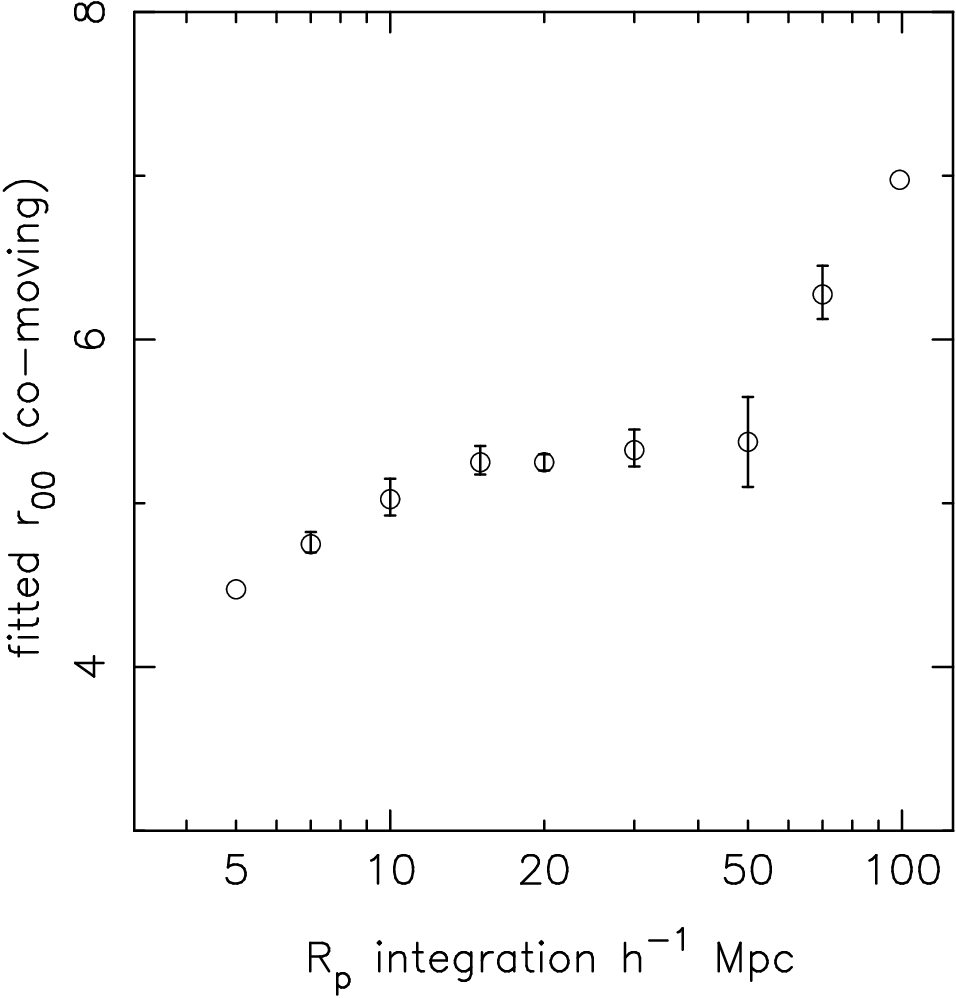} 
\caption{}\end{figure}  
\begin{figure}\figurenum{\nfig}
	\includegraphics[width=\hsize]{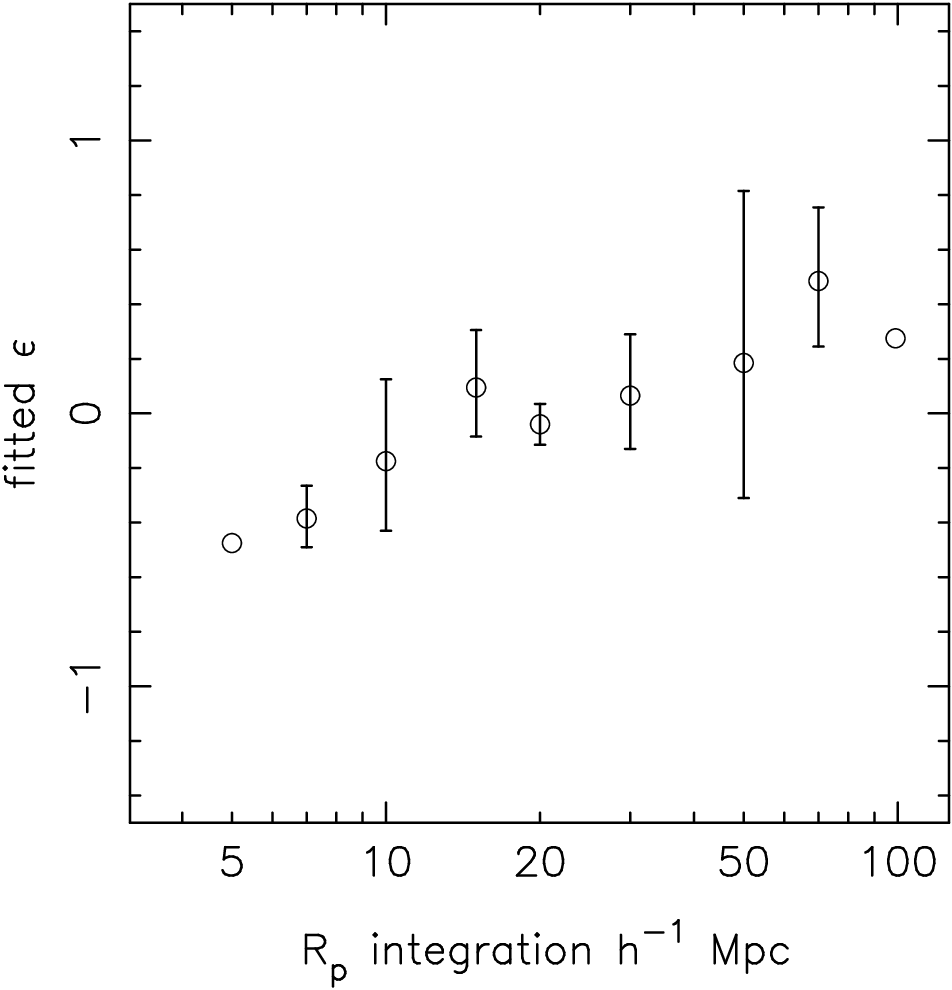} 
\caption{}\end{figure}  

\clearpage

\begin{figure}\figurenum{\nfig}
	\includegraphics[width=\hsize]{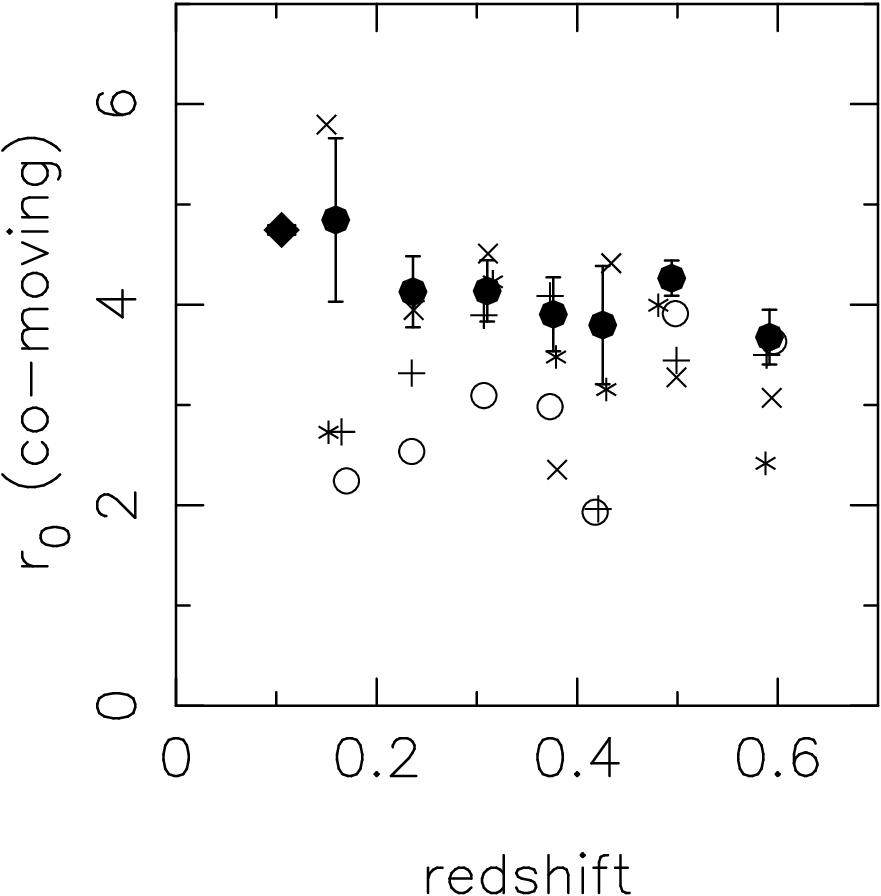} \caption{}\end{figure}  

\clearpage
~
\begin{figure}\figurenum{\nfig}
\begin{picture}(0,0)(0,0)
	\put(0,0){\includegraphics{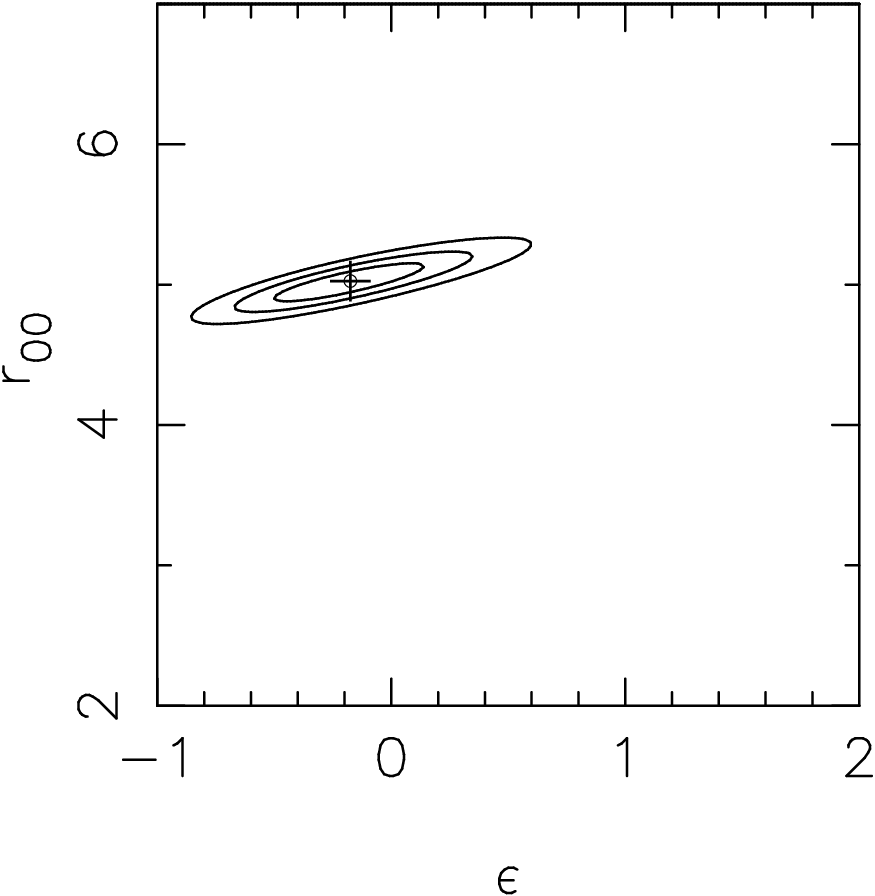}}
	\put(0,0){\includegraphics{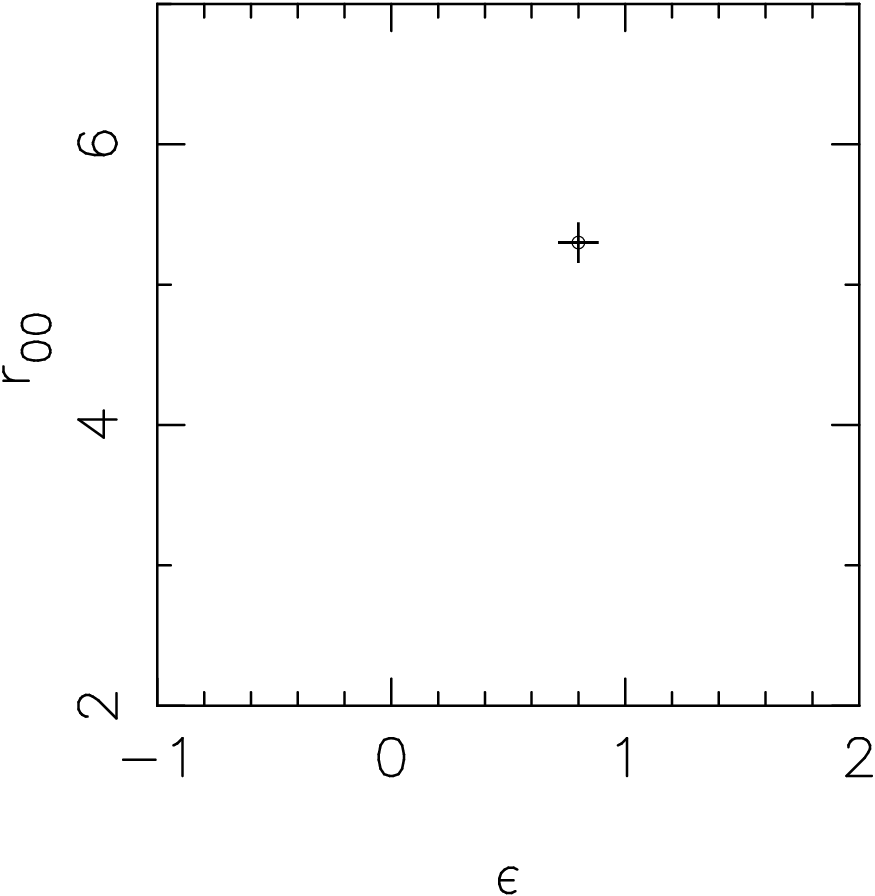}}
	\put(0,0){\includegraphics{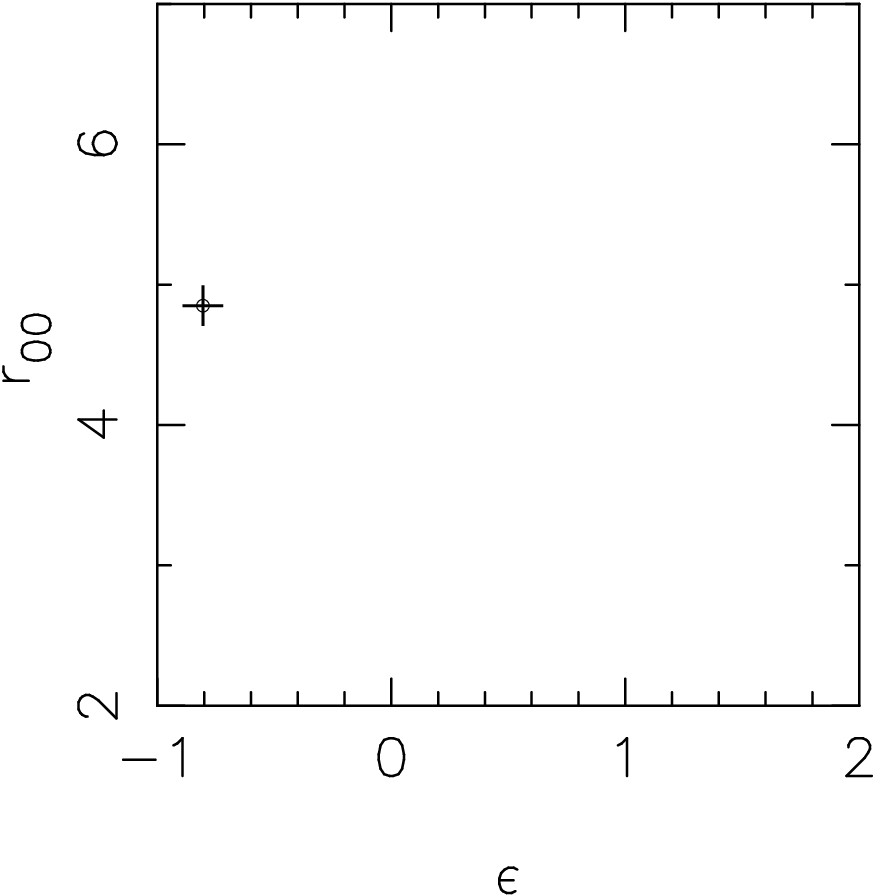}}
	\put(0,0){\includegraphics{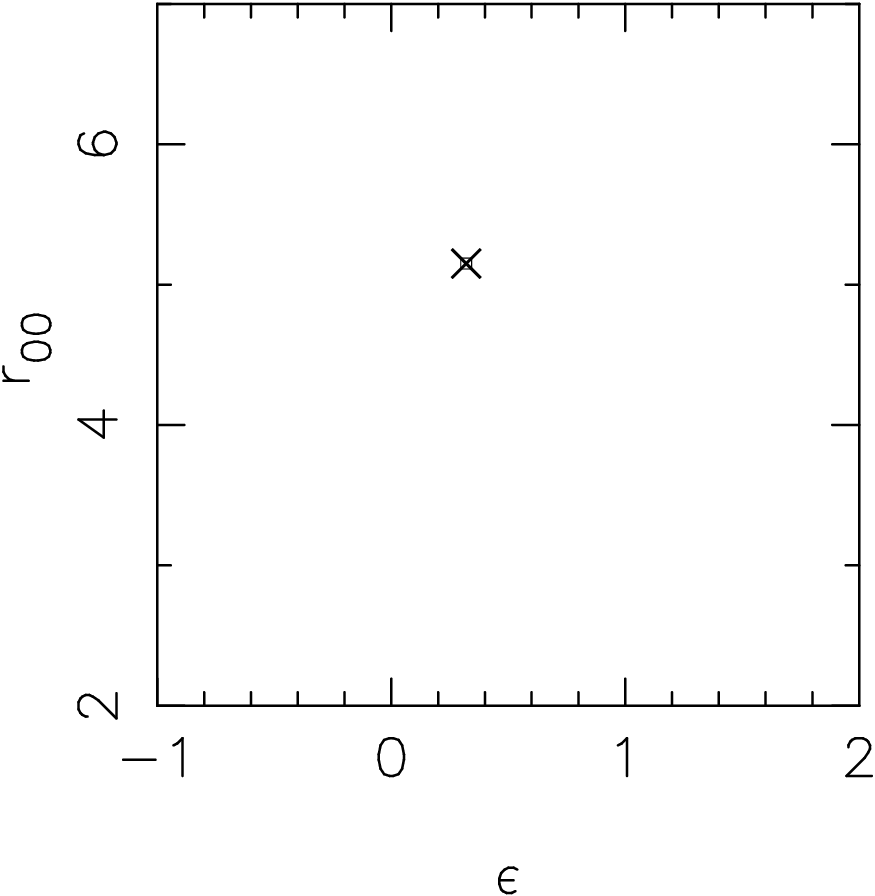}}
	\put(0,0){\includegraphics{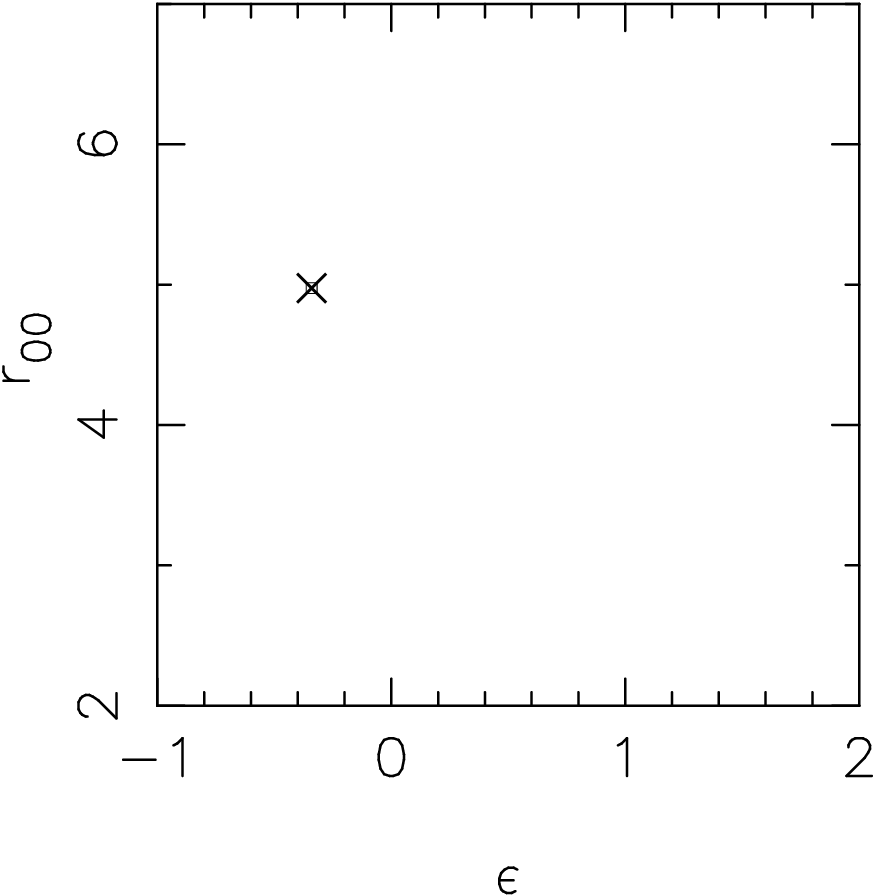}}
\end{picture}
\caption{}\end{figure}  

\clearpage
~
\begin{figure}\figurenum{\nfig}
\begin{picture}(0,0)(0,0)
	\put(0,0){\includegraphics{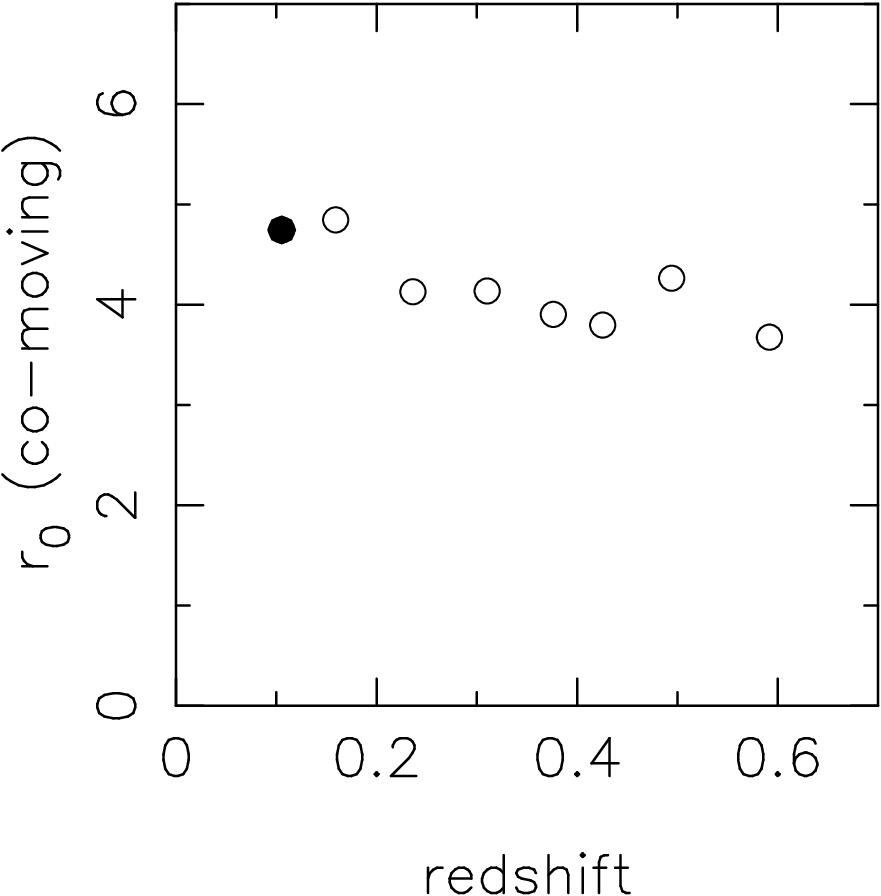}}
	\put(0,0){\includegraphics{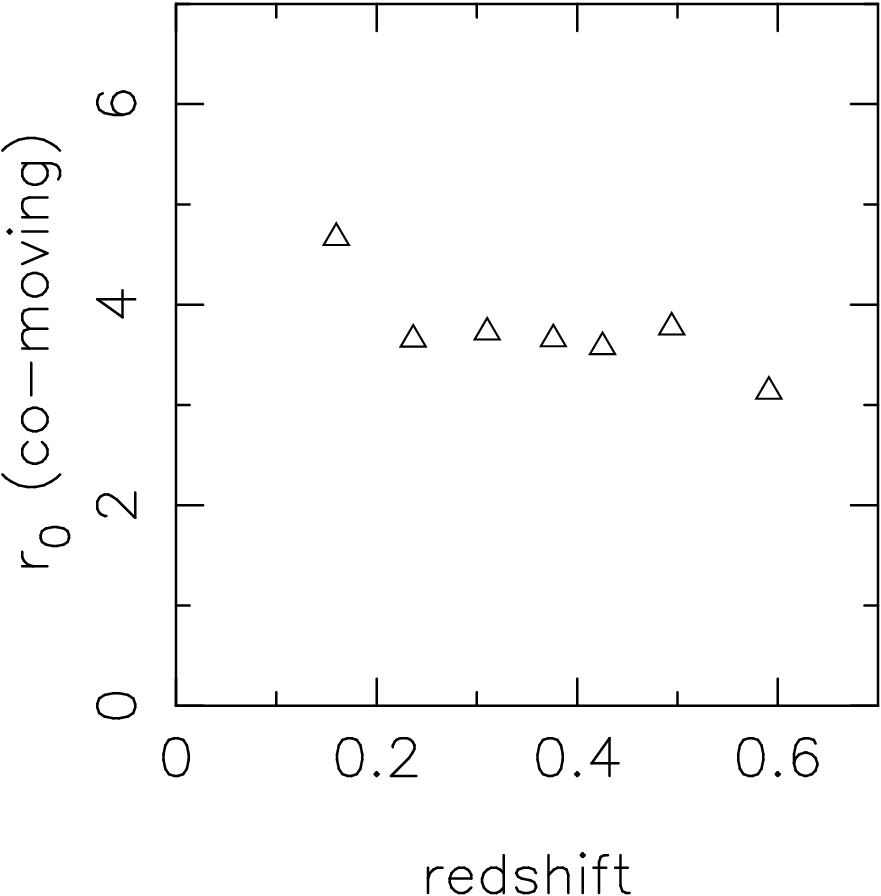}}
	\put(0,0){\includegraphics{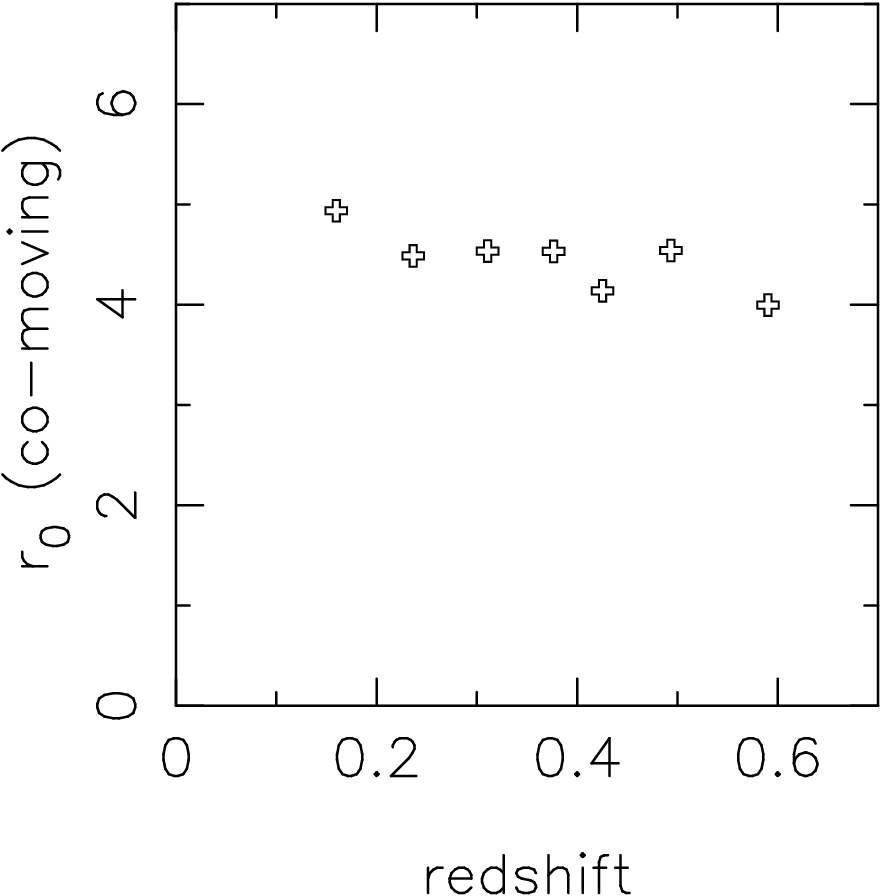}}
\end{picture}
\caption{}\end{figure}  

\end{document}